\documentclass[twocolumn,11pt]{article}

\usepackage[margin=2cm]{geometry}
\usepackage{amsmath,amssymb}
\usepackage{graphicx}
\usepackage[numbers]{natbib}
\usepackage{booktabs}
\usepackage{caption}
\usepackage{float}
\usepackage{microtype}
\usepackage{xcolor}
\usepackage{multirow}
\usepackage{placeins}
\usepackage{stfloats}  
\usepackage[colorlinks=true,linkcolor=blue,citecolor=blue,urlcolor=blue]{hyperref}  

\title{\textbf{Self-Organized Optical Pathways in Optofluidic Photonic Crystals}}

\author{Steven Motta}

\date{March 26, 2026}

\begin{document}
\maketitle

\begin{abstract}
This paper reports FDTD simulations of optofluidic reconfiguration in two-dimensional silicon photonic crystal waveguides, treating structural plasticity (the creation and destruction of optical pathways) via selective fluid infiltration.
Using MPB eigenmode analysis, we decouple bandgap narrowing from defect-mode weakening, showing that defect weakening dominates ($2.4\times$ faster transmission decay than bandgap narrowing at CS$_2$ indices).
Infiltration topology controls signal routing (L-bend selectivity $S = 0.98$), though modulation depth is weak ($\Delta\varepsilon/\varepsilon_\text{Si} = 11\%$).
A phenomenological optothermal feedback model produces self-organized pathways that achieve 63\% of a hand-designed waveguide's bandgap transmission ($7.6\times$ the heavily suppressed empty-crystal baseline).
Amplitude competition between counter-propagating sources produces strong, monotonic pathway steering ($\Delta$COM$_x$ from $+0.03$ to $+4.92\;a$), while pulsed spike-timing-dependent plasticity yields a predictable null result: the timing-sensitive cross-term is suppressed by $>10^2$ when pulse delays exceed the temporal pulse width.
The results provide benchmarks and identify physical limits for bio-inspired reconfigurable optofluidic photonics.
\end{abstract}

\section{Introduction}
\label{sec:intro}

Neuromorphic computing borrows computational principles from biological neural networks~\cite{Schuman2017,deMelo2024}.
Biological neural systems rely on \emph{structural plasticity}, forming new connections, strengthening or weakening existing ones, and pruning unused pathways~\cite{Holtmaat2009}.
Most neuromorphic hardware focuses on synaptic \emph{weight} plasticity; structural plasticity, the dynamic creation and destruction of connectivity itself, has received little attention in photonic systems.

Photonic crystals (PhCs) are well suited to bio-inspired reconfigurable photonics.
A two-dimensional triangular lattice of air holes in a high-index dielectric (e.g., silicon) produces a photonic bandgap: a frequency range where electromagnetic wave propagation is forbidden~\cite{Joannopoulos2008}.
Introducing point or line defects traps and guides photons within the bandgap~\cite{Noda2000}; removing a row of holes creates a W1 waveguide that supports guided modes~\cite{Johnson2000}.
The optical properties of individual holes can be modified \emph{in situ} by infiltrating them with high-refractive-index liquids through microfluidic channels~\cite{Erickson2006,Monat2007}.

\begin{figure*}[t]
  \centering
  \includegraphics[width=\textwidth]{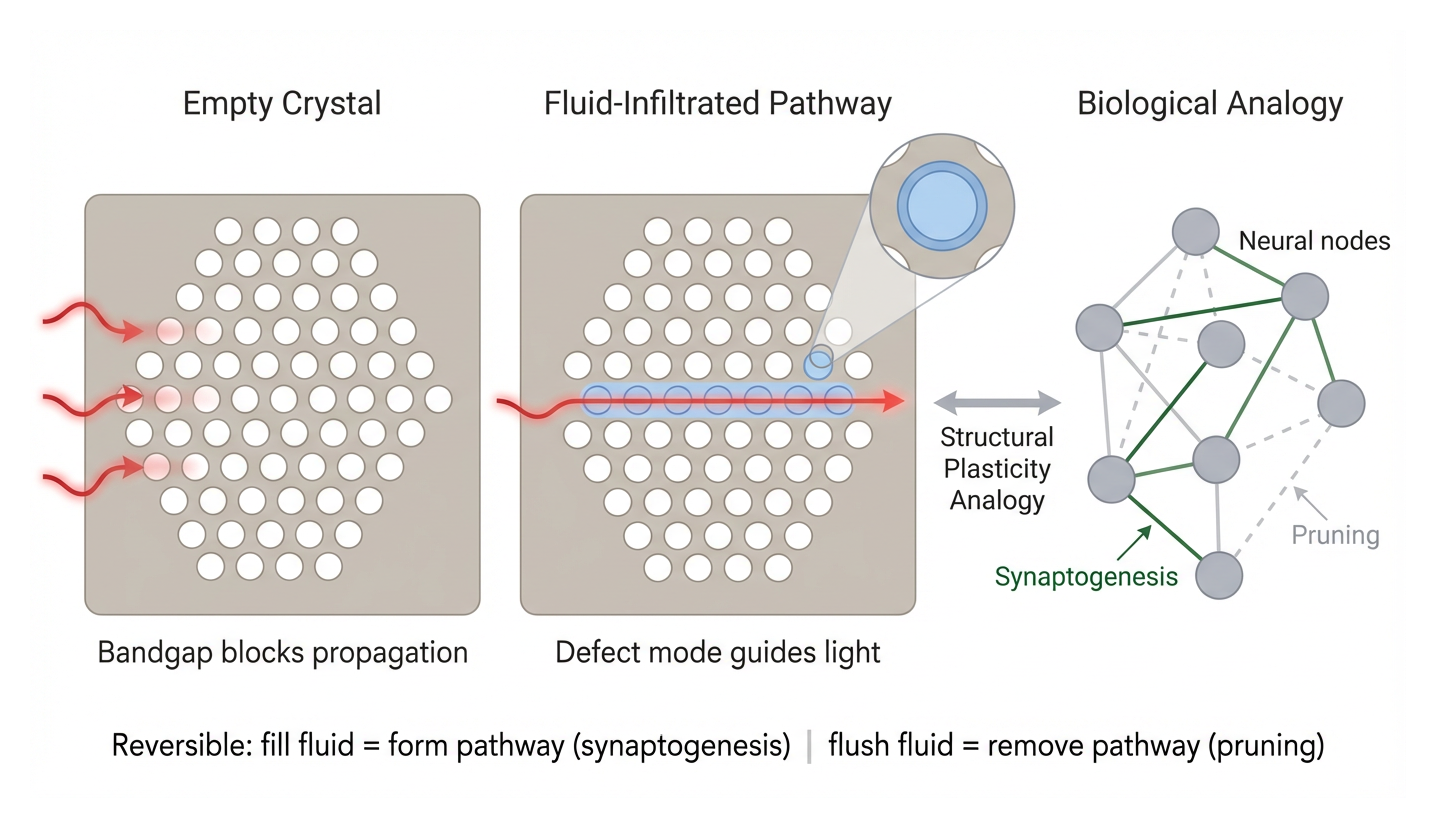}
  \caption{Optofluidic structural plasticity concept. Left: empty photonic crystal, where the bandgap blocks light propagation. Center: fluid-infiltrated pathway; filling a line of holes with CS$_2$ creates a defect mode that guides light. Right: biological analogy; filling/flushing holes corresponds to synaptogenesis/pruning of neural connections. The process is fully reversible.}
  \label{fig:concept}
\end{figure*}

Optofluidic infiltration directly implements structural plasticity: filling a line of holes with fluid creates a refractive-index defect that can guide light (analogous to synaptogenesis), while flushing the fluid restores the bandgap (analogous to synaptic pruning; Fig.~\ref{fig:concept}).
Unlike electronic neuromorphic systems that require physical wiring changes, optofluidic reconfiguration is reversible, reconfigurable on millisecond timescales, and compatible with existing silicon photonic fabrication~\cite{Psaltis2006,Fan2014,Bogaerts2005}.

We use FDTD simulations (MEEP~\cite{Oskooi2010}) and eigenmode analysis (MPB~\cite{Johnson2001}) to systematically investigate this concept.
Our questions:
\begin{enumerate}
  \item Does fluid infiltration of a line of holes create a functional guided-mode waveguide?
  \item How does the transmission depend on the number of infiltrated holes and the fluid refractive index?
  \item Can different infiltration topologies route optical signals to different output ports and implement configurable logic?
  \item Do these effects persist in realistic three-dimensional membrane geometries?
  \item Can optical signals autonomously drive pathway formation through optothermal feedback?
\end{enumerate}

\section{Related Work}
\label{sec:related}

\subsection{Optofluidic Photonic Crystal Reconfiguration}

The concept of tuning photonic crystals via liquid infiltration was proposed theoretically by Busch and John~\cite{Busch1999} and demonstrated experimentally by Erickson et al.~\cite{Erickson2006}, who achieved nanofluidic tuning of SOI photonic crystal circuits with refractive index modulation $\Delta n/n \sim 0.1$ at subwavelength scales.
Smith et al.~\cite{Smith2007} created reconfigurable microfluidic photonic crystal double heterostructure cavities using glass microtip infiltration, achieving quality factors of $Q = 4{,}300$.
Bog et al.~\cite{Bog2008} subsequently demonstrated high-$Q$ microfluidic PhC cavities with $Q = 57{,}000$.
El-Kallassi et al.~\cite{ElKallassi2008} showed that infiltration with photoresponsive liquid crystal blends enables reversible optical tuning; they found that the air-band edge shifts an order of magnitude more than the dielectric-band edge, consistent with our MPB analysis (Sec.~\ref{sec:decouple}).
Intonti et al.~\cite{Intonti2009} demonstrated controlled tuning of PhC cavity resonances over a spectral range exceeding 20~nm with subnanometer accuracy through microinfiltration and subsequent evaporation of water.

These experiments validate the physical mechanism underlying our simulations, though they focused on cavity tuning rather than the waveguide creation and signal routing topology explored here.

\subsection{Alternative Reconfiguration Technologies}

Phase-change materials offer an alternative route to photonic reconfiguration.
Ge$_2$Sb$_2$Te$_5$ (GST) provides non-volatile switching with extinction ratios exceeding 30~dB~\cite{GSTswitch2019} and up to 38 resolvable multi-level optical states~\cite{GSTmulti2022}, and has been used for all-optical neuromorphic networks~\cite{Feldmann2019}.
VO$_2$ achieves ultrafast volatile switching ($\sim$1~ps~\cite{VO2switch2019}) with high extinction ratios ($>$37~dB~\cite{VO2meta2025}).
Electro-optic modulators based on lithium niobate or silicon carrier effects achieve GHz-scale modulation bandwidths~\cite{Reed2010}.

Optofluidic reconfiguration operates on millisecond timescales~\cite{Psaltis2006} with relatively low modulation depth, significantly slower than electronic or phase-change approaches.
Its advantage lies elsewhere: optofluidic systems provide \emph{structural} reconfiguration (creating and destroying entire optical pathways) rather than modulating a fixed pathway, with full reversibility and no material fatigue.
The millisecond timescale is biologically relevant, matching the timescale of structural plasticity in neural circuits~\cite{Holtmaat2009}.

\subsection{Neuromorphic Photonics}

Recent years have seen rapid progress in neuromorphic photonics~\cite{Shastri2021,deMelo2024}.
Tait et al.~\cite{Tait2017} demonstrated silicon photonic weight banks using microring resonator arrays in a broadcast-and-weight architecture, achieving 830~fJ per synaptic operation and establishing the \emph{weight plasticity} paradigm for photonic neuromorphic computing.
Feldmann et al.~\cite{Feldmann2019} demonstrated all-optical spiking neurosynaptic networks using GST phase-change synapses on silicon nitride waveguides, with 4 photonic neurons and 60 non-volatile synapses performing unsupervised pattern recognition. This remains the leading example of phase-change neuromorphic photonics.
Wang et al.~\cite{NeuroPh2024} demonstrated a silicon photonic reservoir computing engine delivering over 200~TOPS (multiply-accumulate operations at GHz bandwidth), reflecting the massive parallelism achievable in integrated photonic circuits, a regime our proof-of-concept study does not yet approach.

Gao et al.~\cite{Gao2023} demonstrated optofluidic memory and self-induced nonlinear optical phase change for reservoir computing, achieving phase changes of 1.7$\pi$~rad via thermocapillary deformation of thin liquid films at kHz switching frequencies. Theirs was the first work directly combining optofluidic mechanisms with neuromorphic computing.

Gao et al.'s platform achieves performance metrics (kHz switching, $1.7\pi$ rad phase shifts) that far exceed our current modulation depth. The potential advantage of our structural-plasticity approach is combinatorial: $N$ holes support $2^N$ distinct topologies, providing an exponentially large configuration space inaccessible to weight-based schemes. For example, in a reconfigurable optical interconnect where $M$ input ports must be dynamically routed to $M$ output ports, weight-based schemes require $M^2$ independently tunable elements, whereas topological reconfiguration of a shared photonic crystal could, in principle, implement arbitrary $M \times M$ permutations by reprogramming a single lattice. Whether this combinatorial advantage compensates for the weak per-hole modulation depth remains undemonstrated.

\begin{figure*}[!b]
  \centering
  \includegraphics[width=\textwidth]{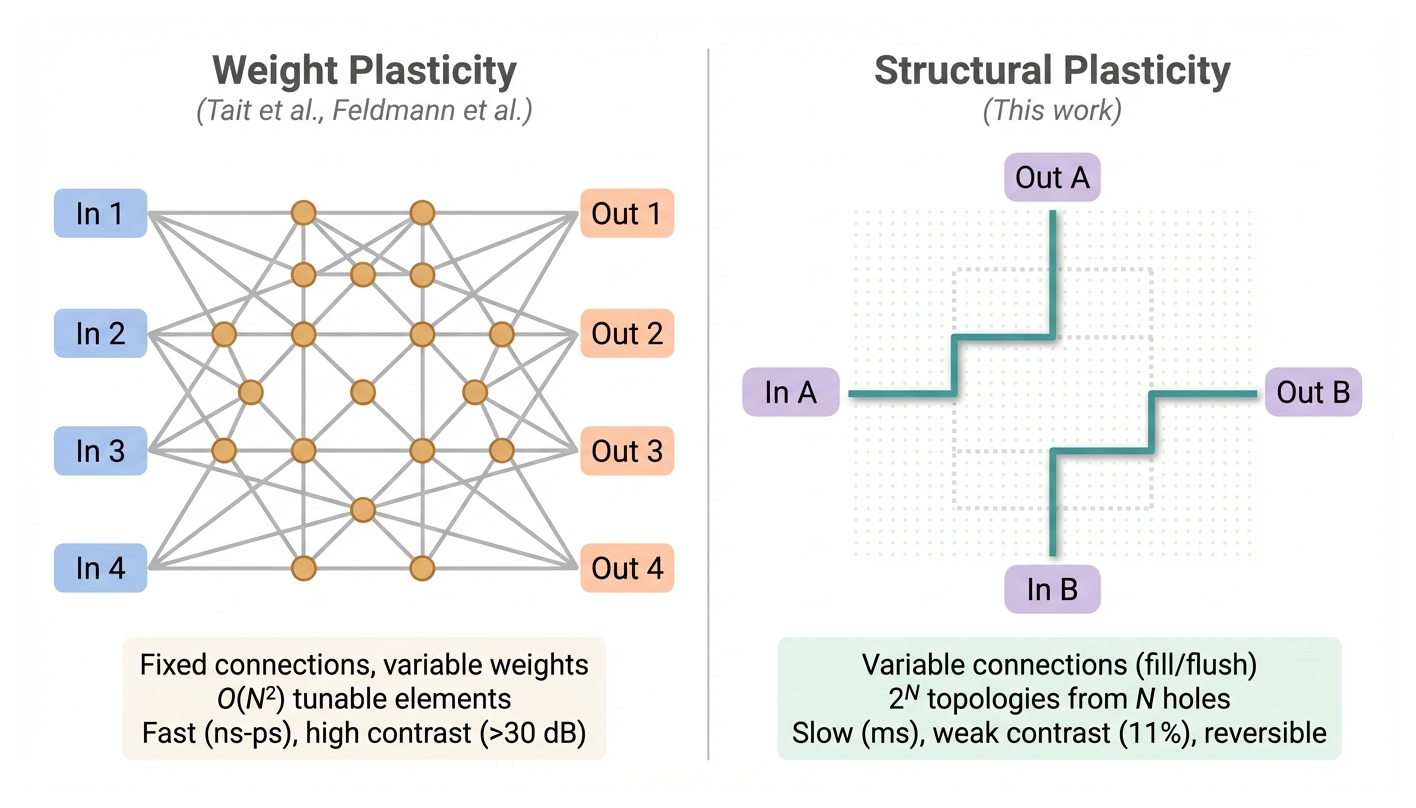}
  \caption{Weight plasticity vs.\ structural plasticity in photonic systems. Left: conventional neuromorphic photonics (Tait, Feldmann) uses fixed waveguide connections with tunable weights ($O(N^2)$ elements for $N \times N$ mapping). Right: our optofluidic approach reconfigures which connections exist by filling/flushing holes in a photonic crystal ($2^N$ possible topologies from $N$ holes). The structural approach trades modulation depth for combinatorial richness.}
  \label{fig:weight_vs_structural}
\end{figure*}

We instead use the photonic crystal lattice itself as the computational substrate, where the \emph{topology} of infiltrated holes determines the connectivity pattern (Fig.~\ref{fig:weight_vs_structural}).
Tait and Feldmann modulate the \emph{strength} of fixed optical connections (weight plasticity); we reconfigure \emph{which} connections exist (structural plasticity).
Our modulation depth is orders of magnitude weaker than these mature platforms.

Optical coincidence detection, required for genuine spike-timing-dependent plasticity, has been demonstrated using two-photon absorption in silicon waveguides~\cite{Liang2005,Xu2007} and $\chi^{(2)}$ sum-frequency generation~\cite{Dayan2004}; Section~\ref{sec:stdp} shows that nonlinear coincidence detection is needed for timing-sensitive plasticity in optofluidic systems.

\section{Methods}
\label{sec:methods}

\subsection{Photonic Crystal Geometry}

We consider a two-dimensional triangular lattice of air holes in silicon with lattice constant $a$ (all lengths normalized to $a$; for telecom wavelengths, $a \approx 400$~nm).
The key geometric parameters are: hole radius $r/a = 0.3$, silicon dielectric constant $\varepsilon_{\text{Si}} = 11.56$ ($n_{\text{Si}} = 3.4$), and slab thickness $h/a = 0.55$ for 3D simulations (corresponding to 220~nm membranes).
The simulation domain uses $N_x = 24$ and $N_y = 17$ lattice periods with perfectly matched layer (PML) absorbing boundaries~\cite{Berenger1994} of thickness $2a$.

\begin{figure*}[t]
  \centering
  \includegraphics[width=\textwidth]{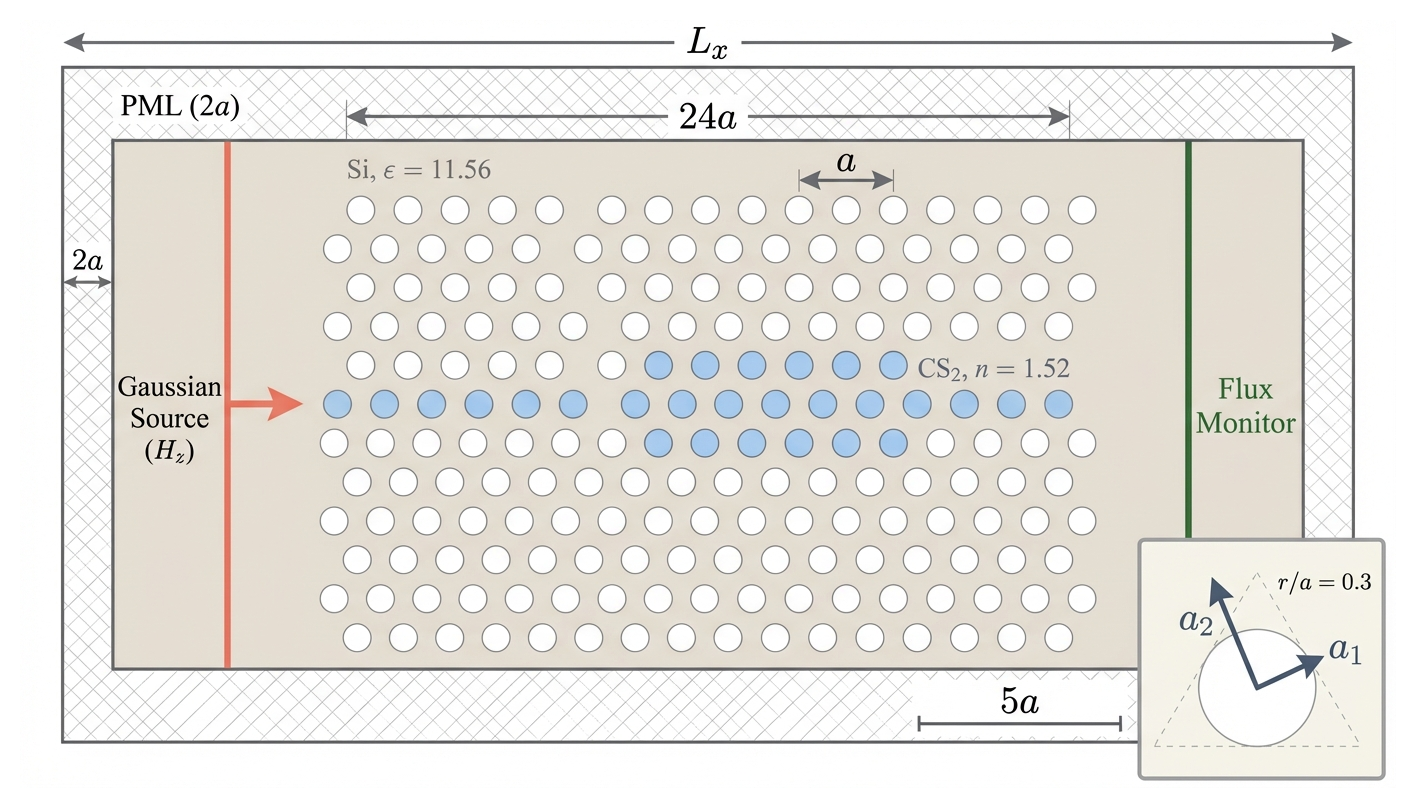}
  \caption{Simulation geometry (top-down view). Triangular lattice of air holes (white) in silicon (gray), with a line of CS$_2$-infiltrated holes (blue) forming the waveguide channel. A Gaussian source (red) launches TE-polarized light; a flux monitor (green) records transmission. PML absorbing boundaries (hatched) surround the domain. Inset: unit cell with $r/a = 0.3$.}
  \label{fig:geometry}
\end{figure*}

For infiltration studies, selected holes are filled with carbon disulfide (CS$_2$, $n = 1.52$, $\varepsilon = 2.31$) unless otherwise specified.
Infiltration configurations are defined as lists of hole indices $(i_x, i_y)$ relative to the crystal center.
Line-length sweeps use odd numbers of holes (1, 3, 5, \ldots, 17) to maintain symmetry about the crystal center; even-numbered lines would break this mirror symmetry, which could introduce additional asymmetric scattering effects.
Figure~\ref{fig:geometry} shows the simulation geometry.

\subsection{Simulation Framework}

All FDTD simulations~\cite{Taflove2005} use MEEP~\cite{Oskooi2010} with subpixel averaging~\cite{Farjadpour2006} enabled.
Band structure calculations use the MIT Photonic Bands (MPB) package~\cite{Johnson2001}.

\paragraph{Sources.}
Broadband transmission measurements use a Gaussian pulse source ($f_{\text{cen}} = 0.25\;c/a$, $\Delta f = 0.1\;c/a$) with $H_z$ polarization (TE modes), placed at $x = -L_x/2 + d_{\text{PML}} + 0.5a$ with transverse extent $1a$.

\paragraph{Monitors.}
Transmission is measured via flux monitors at the opposite domain boundary ($x = +L_x/2 - d_{\text{PML}} - 0.5a$) with transverse size $2a$.
Spectra are resolved across $N_f = 200$ frequency points spanning $f = 0.20$--$0.30\;c/a$.
Simulations terminate when the field amplitude at the monitor decays to $10^{-4}$ of its peak value.

\paragraph{Resolution.}
Production 2D simulations use 40 pixels per lattice constant.
Fast parameter sweeps (disorder, partial infiltration) use 20~pixels/$a$.
3D simulations use 24~pixels/$a$ with a convergence check at 40~pixels/$a$.
A convergence study (Appendix~\ref{app:convergence}) reveals that the production resolution sits on a local maximum of a Cartesian-grid discretization artifact, overestimating integrated bandgap flux by $\sim$30\% relative to the Richardson-extrapolated~\cite{Richardson1911} value. We adopt a conservative $\sim$13\% systematic uncertainty (coefficient of variation across resolutions 32--48~pixels/$a$); relative comparisons between configurations at the same resolution are more robust than absolute values.

\paragraph{Normalization.}
Transmission spectra are normalized to a reference simulation (bulk silicon slab without holes), yielding $T(f) = \Phi_{\text{crystal}}(f) / \Phi_{\text{ref}}(f)$.
This normalization conflates coupling loss (at the crystal boundaries) with propagation physics within the crystal.
Different infiltration topologies may have different coupling efficiencies, which complicates direct comparison of absolute transmission values between configurations.
Relative comparisons within a single topology (e.g., varying line length or fluid index) are less affected.

\section{Results}
\label{sec:results}

\subsection{Band Structure}

\begin{figure*}[t]
  \centering
  \includegraphics[width=\textwidth]{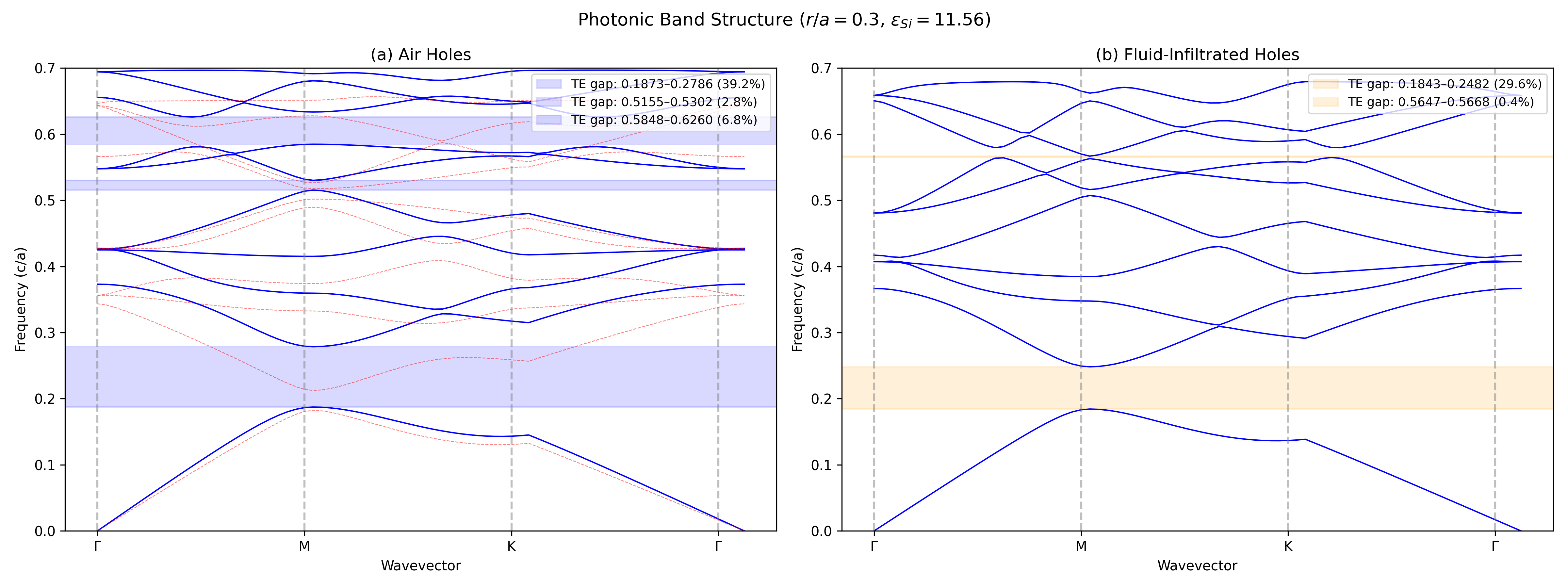}
  \caption{Photonic band structure along high-symmetry directions ($r/a = 0.3$, $\varepsilon_{\text{Si}} = 11.56$). Blue: TE modes; red dashed: TM modes. (a)~Air holes: 39.2\% TE bandgap. (b)~CS$_2$-infiltrated: 30.0\% bandgap.}
  \label{fig:bandstructure}
\end{figure*}

The triangular-lattice photonic crystal exhibits a wide TE bandgap spanning $f = 0.187$--$0.279\;c/a$ with a gap-midgap ratio of 39.2\% (Fig.~\ref{fig:bandstructure}a).
This is consistent with the established value for $r/a = 0.3$ air holes in silicon~\cite{Joannopoulos2008}, validating our simulation setup.
No complete TM bandgap is observed, as expected for this geometry.

When all holes are infiltrated with CS$_2$ ($n = 1.50$, the nearest sweep point to $n = 1.52$), the MPB eigenmode calculation yields a TE bandgap of $f = 0.184$--$0.249\;c/a$ (30.0\% gap-midgap ratio; Fig.~\ref{fig:bandstructure}b).
The lower band edge shifts minimally (from 0.187 to 0.184~$c/a$), while the upper band edge drops significantly (from 0.279 to 0.249~$c/a$), consistent with the asymmetric sensitivity reported by El-Kallassi et al.~\cite{ElKallassi2008}.

\subsection{Waveguide Creation via Infiltration}
\label{sec:waveguide}

\subsubsection{Line-Length Dependence}

\begin{figure*}[!b]
  \centering
  \includegraphics[width=\textwidth]{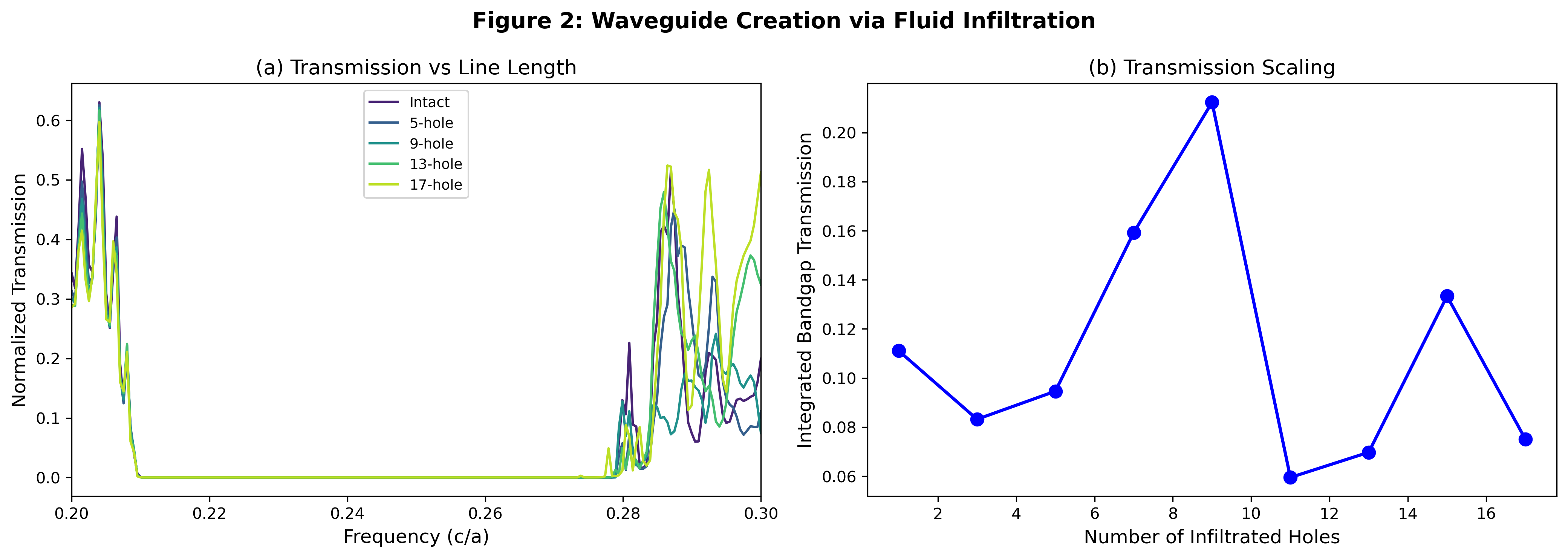}
  \caption{Waveguide creation via fluid infiltration. (a)~Transmission spectra for selected line lengths. (b)~Integrated bandgap transmission vs.\ number of infiltrated holes showing non-monotonic behavior.}
  \label{fig:creation}
\end{figure*}

Figure~\ref{fig:creation} shows the transmission spectra and integrated bandgap transmission as a function of infiltrated line length.
Infiltration does not produce monotonically increasing waveguide-like transmission.
Instead, the integrated bandgap transmission ($\int_{0.22}^{0.28} T(f)\,df$) follows a non-monotonic trajectory with a peak at 9 holes (Table~\ref{tab:linelength}).

\begin{table}[t]
  \centering
  \caption{Integrated bandgap transmission ($f = 0.22$--$0.28\;c/a$) for infiltration line lengths from 1 to 17 holes (CS$_2$, $n = 1.52$). Uncertainties are $\pm 13\%$ from resolution convergence (Appendix~\ref{app:convergence}); adjacent values within these error bars are not statistically distinguishable.}
  \label{tab:linelength}
  \begin{tabular}{@{}lcc@{}}
    \toprule
    Holes & Integrated flux & Regime \\
    \midrule
    1  & $0.111 \pm 0.014$ & Localized defect \\
    3  & $0.083 \pm 0.011$ & Localized defect \\
    5  & $0.095 \pm 0.012$ & Transition \\
    7  & $0.159 \pm 0.021$ & Resonant coupling \\
    9  & $0.212 \pm 0.028$ & Peak resonance \\
    11 & $0.060 \pm 0.008$ & Post-resonance decay \\
    13 & $0.070 \pm 0.009$ & Post-resonance decay \\
    15 & $0.133 \pm 0.017$ & Secondary resonance \\
    17 & $0.075 \pm 0.010$ & Post-resonance decay \\
    \bottomrule
  \end{tabular}
\end{table}

This non-monotonic behavior is qualitatively consistent with coupled-resonator interference~\cite{Yariv1999} in a finite defect chain: short chains support too few resonant modes for efficient transmission, intermediate lengths hit a resonance peak, and longer chains suffer increased scattering loss. We attempted to fit coupled-cavity and Fabry-P\'{e}rot analytical models to the 9 data points but both yield $R^2 < 0$ (worse than a constant mean); these failed fits are documented in Appendix~\ref{app:fabry_perot}. Both models fail, pointing to multi-mode dynamics not captured by single-resonance descriptions.

Given the $\sim$13\% systematic uncertainty from resolution (Sec.~\ref{sec:methods}), several adjacent data points in Table~\ref{tab:linelength} (e.g., 11 and 13~holes at 0.060 and 0.070) are not statistically distinguishable; the secondary peak at 15~holes (0.133) may reflect a genuine Fabry-P\'{e}rot resonance or a Cartesian-grid discretization artifact similar to those identified in Appendix~\ref{app:convergence}.

The non-monotonic behavior differs from W1 waveguides, where removing holes produces near-unity in-gap transmission.
The fundamental limitation is the weakness of the refractive-index perturbation: $\Delta\varepsilon / \varepsilon_{\text{Si}} = 1.31/11.56 = 11\%$.

\subsection{Decoupling Bandgap Narrowing from Defect Weakening}
\label{sec:decouple}

\begin{figure}[t]
  \centering
  \includegraphics[width=\columnwidth]{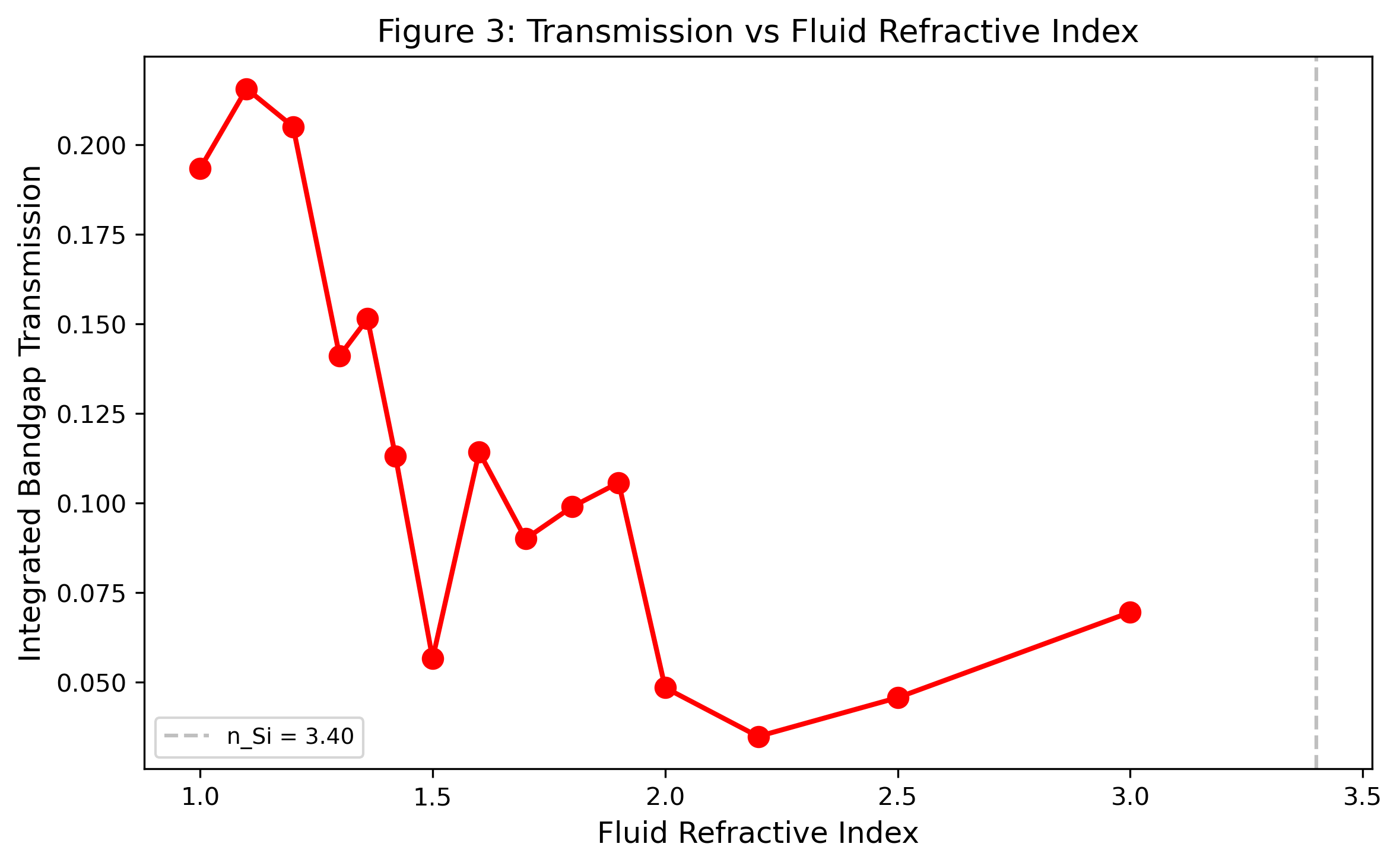}
  \caption{Integrated bandgap transmission vs.\ fluid refractive index for an 11-hole infiltrated line. Transmission decreases with increasing fluid index above $n \approx 1.1$.}
  \label{fig:fluid}
\end{figure}

The counterintuitive decrease in transmission with increasing fluid index (Fig.~\ref{fig:fluid}) reflects two competing mechanisms: (1)~bandgap narrowing allows more transmission at edge frequencies, and (2)~defect weakening reduces the perturbation that supports the defect mode.
To decouple these effects, we performed an MPB eigenmode sweep across 15 fluid refractive indices ($n = 1.0$--$3.0$), computing the TE bandgap independently of any defect structure.

\begin{figure*}[t]
  \centering
  \includegraphics[width=\textwidth]{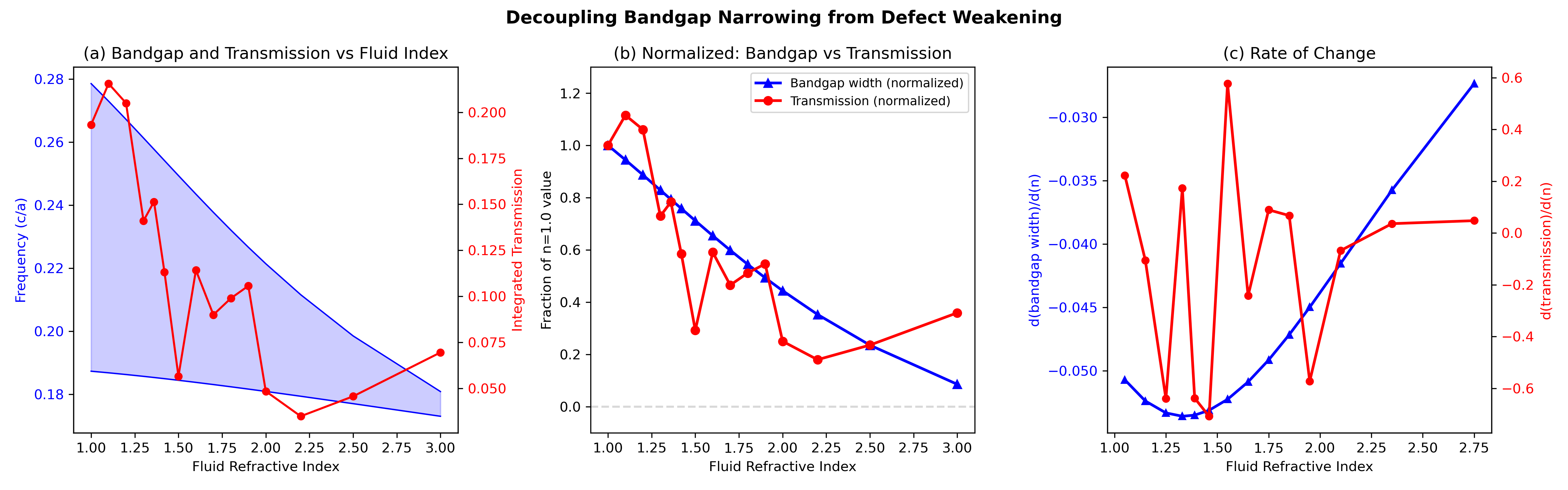}
  \caption{Decoupling bandgap narrowing from defect weakening. (a)~Bandgap edges (blue shading) and integrated transmission (red) vs.\ fluid index. (b)~Normalized comparison: both quantities relative to their $n = 1.0$ values. Transmission drops $2.4\times$ faster than bandgap width, proving defect weakening dominates. (c)~Rate of change $d/dn$ for both quantities.}
  \label{fig:decouple}
\end{figure*}

Figure~\ref{fig:decouple} shows the quantitative comparison.
At $n = 1.52$ (CS$_2$), the bandgap width retains 71.2\% of its air-hole value (narrowing from 0.091 to 0.065~$c/a$), while the FDTD transmission retains only 29.2\% of the $n = 1.0$ value.
This 2.4$\times$ faster decay of transmission relative to bandgap width shows that \textbf{defect weakening dominates over bandgap narrowing}.

As the fluid index increases, the infiltrated holes become more similar to the silicon background, reducing the effective defect perturbation.
At $n = 3.0$, the bandgap retains only 8.5\% of its original width (gap-midgap ratio drops from 39.2\% to 4.4\%), while the defect has become so weak that the remaining transmission is primarily edge leakage.

\subsection{Dynamic Signal Rerouting}
\label{sec:rerouting}

\begin{figure*}[!b]
  \centering
  \includegraphics[width=\textwidth]{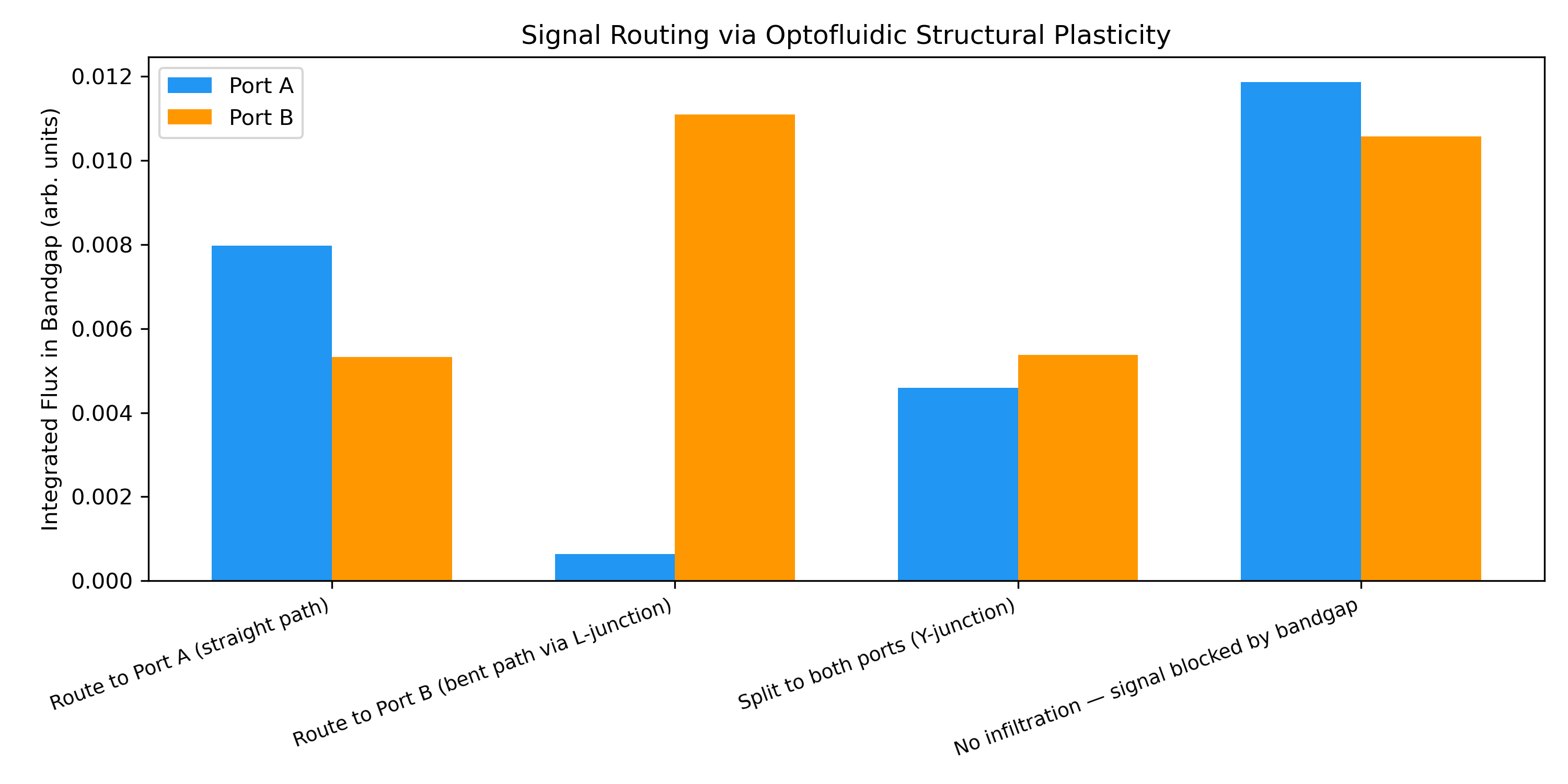}
  \caption{Signal routing via infiltration topology. Integrated bandgap flux at Port~A and Port~B for four configurations. The L-bend configuration achieves $S = 0.89$ directional selectivity.}
  \label{fig:rerouting}
\end{figure*}

The rerouting demonstration (Fig.~\ref{fig:rerouting}) tests whether infiltration topology can direct optical signals to different output ports.
Four configurations are compared: straight path to Port~A, L-bend to Port~B, Y-junction split, and no infiltration (blocked).

Table~\ref{tab:rerouting} summarizes the results.
The key metric is the \emph{directional selectivity} $S = |F_A - F_B| / (F_A + F_B)$, where $F_A$ and $F_B$ are the integrated fluxes at each port.

\begin{table}[t]
  \centering
  \caption{Rerouting results: integrated bandgap flux at each port, raw directional selectivity $S_{\text{raw}}$, and background-corrected selectivity $S_{\text{corr}}$ (see text). The blocked configuration serves as the scattered-light baseline.}
  \label{tab:rerouting}
  \begin{tabular}{@{}lcccc@{}}
    \toprule
    Configuration & Port A & Port B & $S_{\text{raw}}$ & $S_{\text{corr}}$ \\
    \midrule
    Route A (straight) & 0.0080 & 0.0053 & 0.20 & 0.29 \\
    Route B (L-bend)   & 0.0006 & 0.0111 & 0.89 & 0.98 \\
    Y-split            & 0.0046 & 0.0054 & 0.08 & 0.06 \\
    Blocked (baseline) & 0.0119 & 0.0106 & 0.06 & --- \\
    \bottomrule
  \end{tabular}
\end{table}

We report both raw and background-corrected selectivity.
For background correction, we subtract a uniform scattered-light floor $F_{\text{bg}} = (F_{A,\text{blocked}} + F_{B,\text{blocked}})/2 = 0.0112$ from each port, then recompute $S_{\text{corr}} = |F'_A - F'_B|/(|F'_A| + |F'_B|)$.
This correction is imperfect (the blocked flux is not a simple additive background, since infiltrated holes modify the scattering geometry), but it provides a bound on the true routing selectivity.

The Route~B configuration achieves $S_{\text{corr}} = 0.98$, confirming strong topological routing.
Route~A shows $S_{\text{corr}} = 0.29$, weakly directional at best.
Route~A's raw flux at both ports is \emph{below} the blocked baseline, meaning the straight infiltration path may actually suppress transmission relative to the unperturbed crystal, possibly by disrupting scattering paths that happen to direct flux toward Port~A in the blocked geometry.
The Y-split shows near-zero selectivity ($S_{\text{corr}} = 0.06$) consistent with symmetric splitting.

Only the L-bend configuration demonstrates unambiguous directional routing.
The straight-through configuration's weak performance likely reflects the inherently low index perturbation: the L-bend benefits from a longer path through the crystal and a $90^\circ$ change in propagation direction that suppresses the direct scattered-light background at Port~B.

\subsection{Configurable Input-Output Mappings}
\label{sec:logic}

\begin{figure*}[t]
  \centering
  \includegraphics[width=\textwidth]{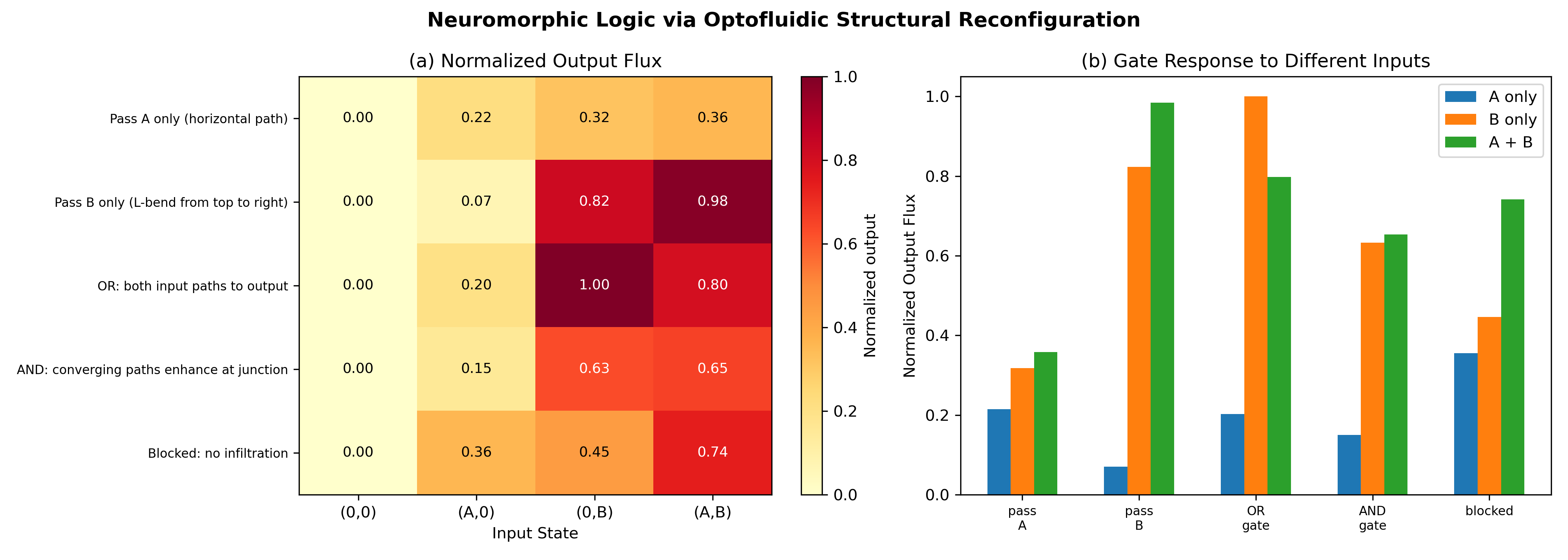}
  \caption{Configurable input-output mappings via infiltration topology. (a)~Normalized output flux for five topologies and four input states. (b)~Comparison of topology responses. The ``blocked'' configuration transmits non-negligible scattered flux, limiting the effective signal-to-background ratio.}
  \label{fig:logic}
\end{figure*}

To explore whether infiltration topology can implement distinct input-output transfer functions, we configured five infiltration patterns that route signals from two input ports (Source~A: horizontal, Source~B: vertical) to a single output port (Fig.~\ref{fig:logic}).
The five topologies are: (1)~horizontal path (``Pass~A''), (2)~L-bend path (``Pass~B''), (3)~both paths (``Dual-path''), (4)~converging paths with widened junction (``Junction''), and (5)~no infiltration (``Blocked'').

\textbf{Caveat}: as shown below, several topology/input combinations produce output flux below the scattered-light floor of the blocked baseline, limiting the conclusions that can be drawn about guided-mode routing. We present the full data for completeness but restrict our claims to configurations with signal-to-background ratio exceeding 1.5$\times$.

For each topology, we run simulations with all four input combinations: (0,0), (A,0), (0,B), and (A,B).
Table~\ref{tab:logic} reports the raw flux values and key diagnostic ratios.

\begin{table}[t]
  \centering
  \caption{Input-output mapping results. $F_{A}$, $F_B$, $F_{AB}$: output flux for A-only, B-only, and both-inputs-active states. $F_{AB}/(F_A + F_B)$: superposition ratio (1.0 = linear, $>$1 = cooperative, $<$1 = destructive interference). S/B: signal-to-background ratio vs.\ blocked configuration.}
  \label{tab:logic}
  \begin{tabular}{@{}lcccc@{}}
    \toprule
    Topology & $F_A$ & $F_B$ & $F_{AB}$ & $\frac{F_{AB}}{F_A+F_B}$ \\
    \midrule
    Pass A    & 0.012 & 0.018 & 0.020 & 0.67 \\
    Pass B    & 0.004 & 0.046 & 0.055 & 1.11 \\
    Dual-path & 0.011 & 0.056 & 0.045 & 0.66 \\
    Junction  & 0.008 & 0.036 & 0.037 & 0.83 \\
    Blocked   & 0.020 & 0.025 & 0.042 & 0.93 \\
    \bottomrule
  \end{tabular}
\end{table}

\paragraph{Superposition test.}
A true AND gate requires the combined output $F_{AB}$ to exceed the linear superposition $F_A + F_B$ (cooperative enhancement).
A true OR gate requires $F_{AB} \geq \max(F_A, F_B)$.
Testing these conditions: the ``Junction'' topology gives $F_{AB}/(F_A + F_B) = 0.83$ (sub-linear, not AND-like); the ``Dual-path'' gives $F_{AB} = 0.045 < F_B = 0.056$ (destructive interference, not OR-like).
Only the ``Pass~B'' topology shows weakly super-linear behavior ($F_{AB}/(F_A + F_B) = 1.11$), but this is within the systematic uncertainty.
We therefore do not claim that these topologies implement Boolean logic; rather, they demonstrate \emph{distinct input-output transfer functions} that differ qualitatively between topologies.

\paragraph{Signal-to-background.}
The blocked configuration transmits significant scattered flux (0.020--0.042), which exceeds several active-topology outputs.
For example, Pass~A's $F_A = 0.012$ is only $0.6\times$ the blocked $F_A = 0.020$, below the scattered-light floor.
The configurations with the best signal-to-background are Pass~B ($F_B/F_{B,\text{blocked}} = 1.8\times$) and Dual-path ($F_B/F_{B,\text{blocked}} = 2.2\times$).
The limited contrast stems from the same weak perturbation that limits waveguide transmission throughout this study.

\paragraph{Topology-dependent transfer functions.}
Despite the S/B limitations, the five topologies produce measurably different transfer functions.
The Pass~B $F_B/F_A$ ratio (11.7) differs dramatically from Pass~A ($F_B/F_A = 1.5$) and Blocked ($F_B/F_A = 1.3$), confirming that topology controls which input port couples most effectively to the output.
Changing the infiltration pattern changes the computational function, even when the absolute modulation depth is small.

\subsection{Three-Dimensional Validation}

\begin{figure*}[!b]
  \centering
  \includegraphics[width=\textwidth]{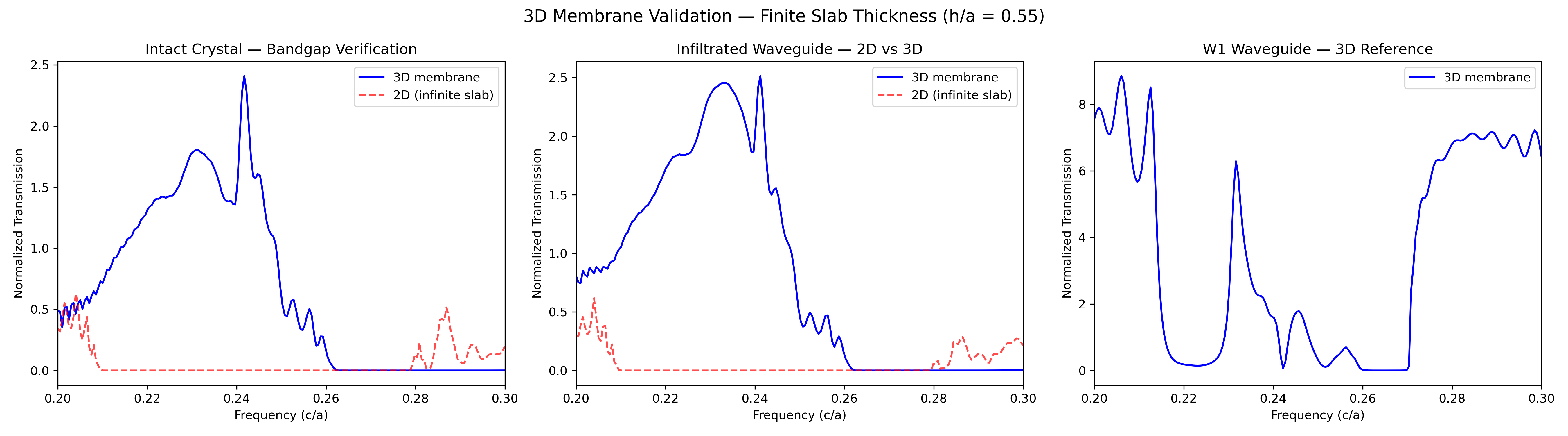}
  \caption{3D membrane validation ($h/a = 0.55$). Transmission spectra for intact crystal and infiltrated waveguide. The bandgap and infiltration effects persist in realistic 3D geometries.}
  \label{fig:3d}
\end{figure*}

Full 3D membrane simulations ($h/a = 0.55$, resolution 24~pixels/$a$) were run for 5 configurations plus a convergence check at 40~pixels/$a$ (Table~\ref{tab:3d}).
The photonic bandgap persists in 3D (Fig.~\ref{fig:3d}), with the bandgap region shifting slightly due to the light-cone cutoff in slab waveguides.

\begin{table}[t]
  \centering
  \caption{3D membrane validation: total integrated spectral flux for each configuration (resolution 24~pixels/$a$, $h/a = 0.55$).}
  \label{tab:3d}
  \begin{tabular}{@{}lc@{}}
    \toprule
    Configuration & Integrated flux \\
    \midrule
    Reference (no holes) & 32.93 \\
    Empty crystal (3D) & 25.58 \\
    Line-infiltrated (3D) & 27.48 \\
    W1 waveguide (3D) & 33.47 \\
    Route B, Port A & 26.56 \\
    Route B, Port B & 3.50 \\
    \bottomrule
  \end{tabular}
\end{table}

The key qualitative features persist: the empty crystal blocks most transmission (flux reduced to 78\% of reference), the W1 waveguide restores near-full transmission (102\% of reference, with the slight excess likely from mode conversion effects), and line infiltration produces only a modest $1.07\times$ increase over the empty crystal, within the $\sim$13\% systematic uncertainty. The infiltration effect therefore cannot be confirmed to persist in 3D at this resolution. The directional rerouting asymmetry, by contrast, is well above the noise floor.
The Route~B configuration shows strong directional asymmetry ($\text{Port B}/\text{Port A} = 0.13$), confirming that rerouting effects persist in 3D.
A convergence check at 40~pixels/$a$ shows 10\% agreement with the res-24 result, which is the same order as our systematic uncertainty and therefore does not provide an independent constraint on accuracy.

\section{Self-Organized Pathway Formation via Optothermal Feedback}
\label{sec:hebbian}

Sections~\ref{sec:results}--\ref{sec:rerouting} covered externally commanded reconfiguration, where an operator selects which holes to infiltrate.
Given a feedback model in which optical absorption drives local heating and thermocapillary-driven fluid infiltration, what pathway topologies emerge? The update rule is phenomenological and Hebbian by construction~\cite{Hebb1949} (``pathways that carry light attract more fluid''); the question is not \emph{whether} self-organization occurs (it is built into the equations) but \emph{what} structures the model produces and whether they are functionally useful.

\subsection{Feedback Model}

\begin{figure}[t]
  \centering
  \includegraphics[width=\columnwidth]{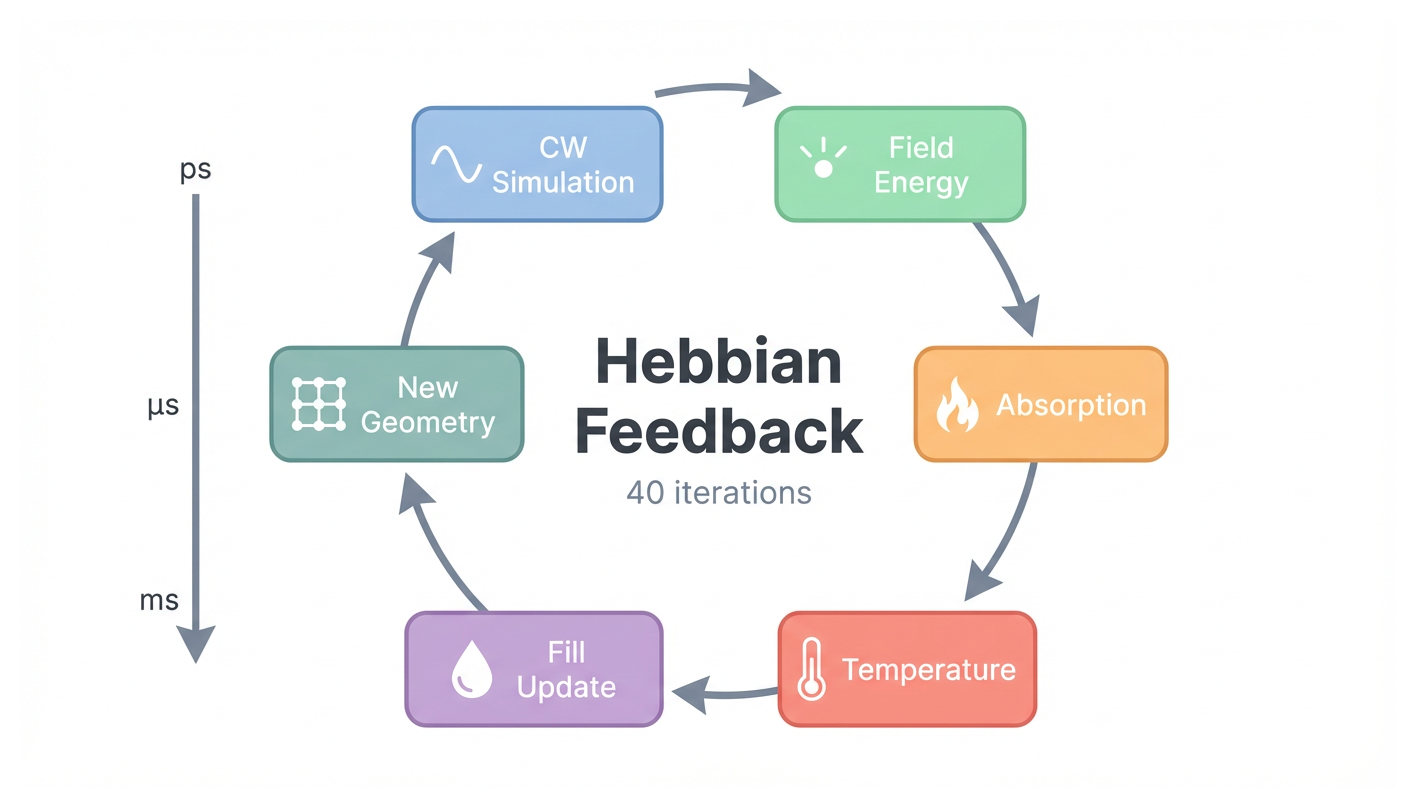}
  \caption{The optothermal Hebbian feedback loop. Six steps cycle iteratively: (1)~MEEP CW electromagnetic solve, (2)~field energy extraction at each hole, (3)~absorbed power computation with $\eta_\text{seed}$ floor, (4)~normalized temperature map with thermal diffusion, (5)~fill fraction update (hot holes fill, cold holes drain), (6)~geometry update with new dielectric values. The cycle repeats for 40~iterations, spanning optical (ps), thermal ($\mu$s), and fluidic (ms) timescales.}
  \label{fig:hebbian_loop}
\end{figure}

We implement a split-timescale iterative loop on a reduced $16 \times 11$ triangular lattice (204~holes, resolution 20~pixels/$a$).
Each iteration consists of:
\begin{enumerate}
  \item \textbf{MEEP CW solve}: run for 300 time units with a ContinuousSource at $f = 0.25\;c/a$.
  \item \textbf{Field energy extraction}: compute $U_i = \frac{1}{2}\varepsilon\int|\mathbf{E}|^2\,dA$ over a bounding box around each hole $i$ using \texttt{electric\_energy\_in\_box}.
  \item \textbf{Absorbed power}: $P_{\text{abs},i} = \omega \cdot \max(\eta_i, \eta_{\text{seed}}) \cdot \text{Im}(\varepsilon) \cdot 2U_i / \varepsilon_{\text{eff},i}$, where $\eta_{\text{seed}} = 0.01$ is a phenomenological seeding parameter.
  The parameter $\eta_{\text{seed}}$ is a computational convenience that bootstraps the feedback by allowing nominally empty holes to absorb weakly; without it ($\eta_{\text{seed}} = 0$), the feedback cannot propagate beyond the initially seeded layer and the simulation deadlocks. This deadlock at $\eta_{\text{seed}} = 0$ is a limitation of the phenomenological model, not necessarily a statement about the physics: in reality, thermal conduction from neighboring hot holes would warm adjacent empty holes (our thermal diffusion term partially captures this), and trace fluid wetting along channel walls would provide nonzero initial absorption. A more rigorous approach would replace the global $\eta_{\text{seed}}$ with a spatially local thermal conduction kernel; we use the simpler global parameter for tractability.
  Results are robust to order-of-magnitude variation in $\eta_{\text{seed}}$ ($0.001$--$0.1$; see Sec.~\ref{sec:hebbian_sensitivity}).
  \item \textbf{Normalized $\Delta T$}: normalize absorbed powers to $[0,1]$ and apply thermal diffusion (15\% of nearest-neighbor average $\Delta T$).
  \item \textbf{Fill fraction update}: $\Delta\eta_i = \gamma \cdot \Delta T_i \cdot (1 - \eta_i) - r_{\text{drain}} \cdot \eta_i / (1 + \Delta T_i / \Delta T_{\text{thresh}})$, where $\gamma = 0.8$ is the feedback gain, $r_{\text{drain}} = 0.05$ is the passive drain rate, and $\Delta T_{\text{thresh}}$ is the median of all nonzero $\Delta T_i$ values at each iteration (an adaptive threshold that separates ``active'' from ``inactive'' holes).
  \item \textbf{Geometry update}: $\varepsilon_{\text{eff},i} = (1 - \eta_i) + \eta_i \cdot \varepsilon_{\text{fluid}}$ via linear effective medium mixing~\cite{Aspnes1982}.
\end{enumerate}
The absorbing fluid is modeled as CS$_2$ doped with an absorbing dye ($\text{Im}(\varepsilon) = 0.05$ by default), using MEEP's \texttt{D\_conductivity} parameter.
Pure CS$_2$ has negligible absorption at telecom wavelengths; the imaginary part represents an intentional dopant (e.g., a near-IR absorbing dye) added to enable the optothermal feedback mechanism.
The full feedback cycle is illustrated in Fig.~\ref{fig:hebbian_loop}.
The feedback simulations use resolution 20~pixels/$a$ for computational tractability (40 FDTD solves per scenario); given the $\sim$13\% systematic uncertainty at res$=$40 (Appendix~\ref{app:convergence}), absolute flux values at res$=$20 carry larger uncertainty, but the qualitative pathway topology should be robust since all iterations use the same resolution.

\subsection{Three Scenarios}

\begin{figure*}[t]
  \centering
  \includegraphics[width=\textwidth]{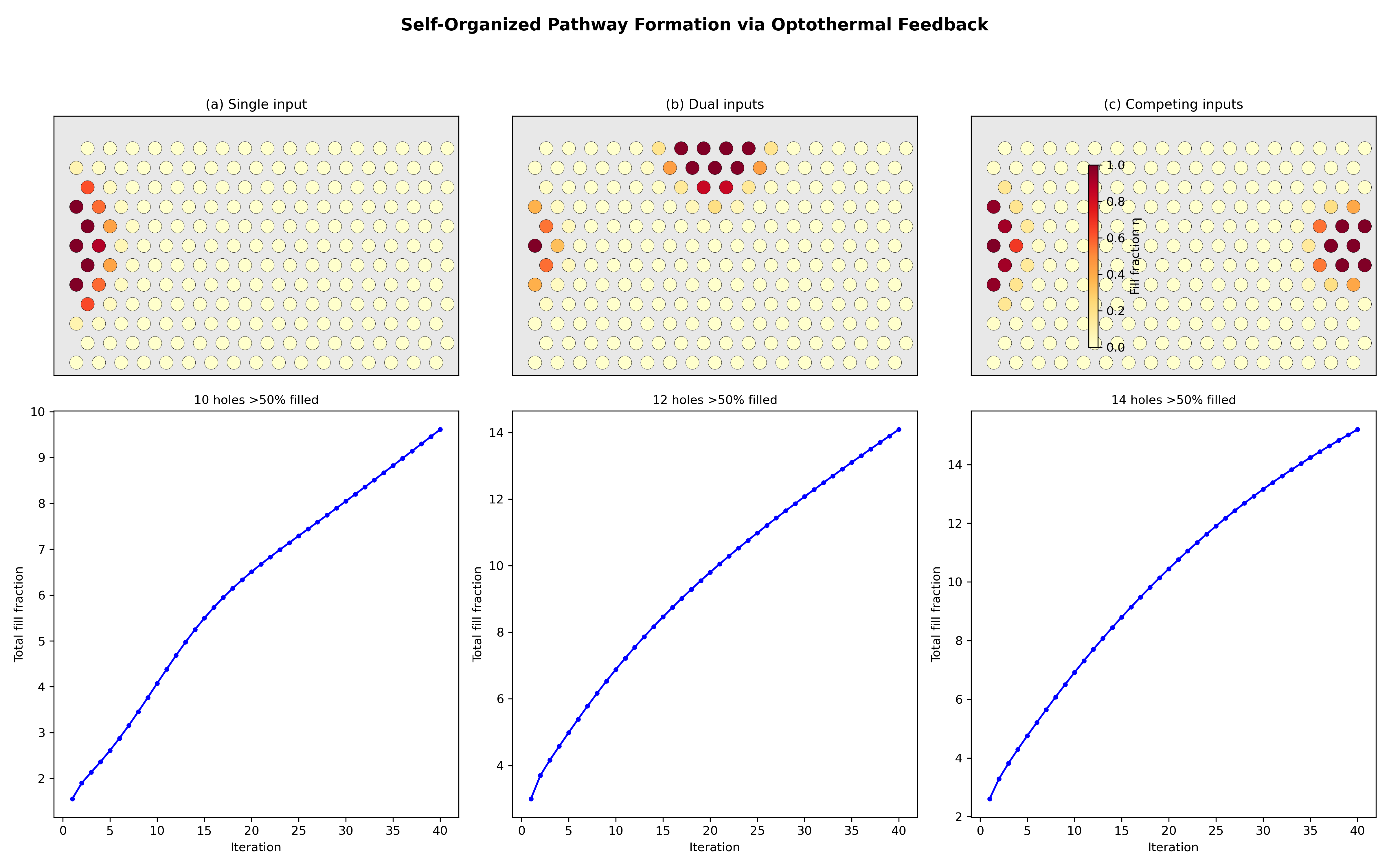}
  \caption{Self-organized pathway formation via optothermal feedback. Top row: final fill-fraction maps ($\eta$, color scale) after 40~iterations for three source configurations. Bottom row: total fill fraction convergence. (a)~Single input: 10~holes $>$50\% filled. (b)~Dual inputs: 12~holes $>$50\% filled, two distinct pathways. (c)~Competing inputs: 14~holes $>$50\% filled, symmetric infiltration.}
  \label{fig:hebbian}
\end{figure*}

\begin{figure*}[!b]
  \centering
  \includegraphics[width=\textwidth]{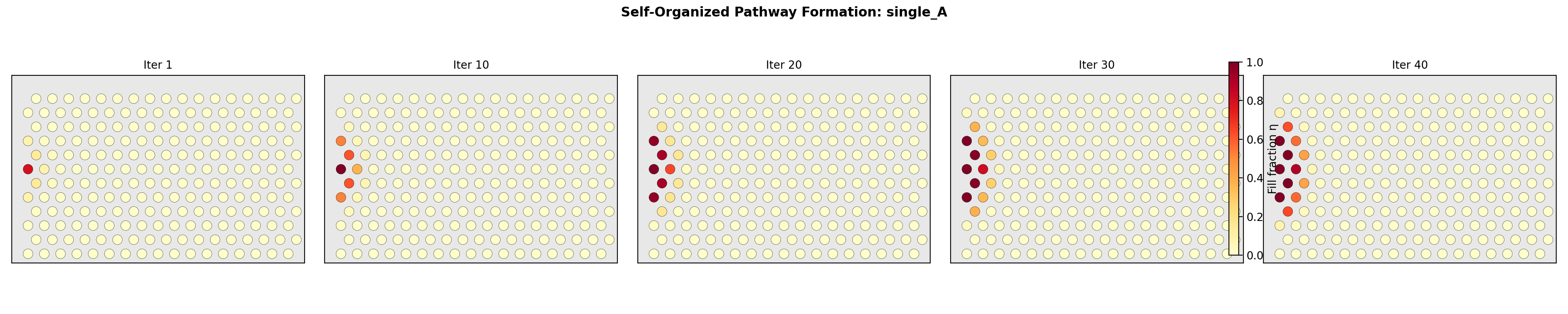}
  \caption{Time-lapse of pathway self-organization for the single-input scenario at iterations 1, 10, 20, 30, and 40. The pathway progressively extends from the source (left edge) into the crystal, with initially seeded holes reaching full saturation while neighboring holes are activated through thermal diffusion.}
  \label{fig:timelapse}
\end{figure*}

We test three source configurations (Fig.~\ref{fig:hebbian}):

\paragraph{Single input.}
A single source on the left face drives pathway formation into the crystal.
After 40~iterations, 12~holes exceed $\eta > 0.1$ and 10~holes exceed $\eta > 0.5$, forming a localized pathway extending $\sim$3 lattice constants from the source.
The total fill fraction $\Sigma\eta$ grows monotonically from 1.6 to 9.6, with the active hole count increasing from 3 to 12 as thermal diffusion activates neighboring holes.
The convergence curves (Fig.~\ref{fig:hebbian}, bottom row) show $\Sigma\eta$ approaching a plateau with the per-iteration increment $\Delta\Sigma\eta$ declining from $\sim$0.3 to $\sim$0.15 over the final 10~iterations; the maximum per-hole change $\max|\Delta\eta|$ decreases from 0.07 to 0.025, approaching but not reaching the convergence threshold of 0.003.
Extrapolation suggests the pathway topology is established by iteration $\sim$20 with subsequent iterations deepening existing holes rather than activating new ones (Fig.~\ref{fig:timelapse}).

\paragraph{Dual inputs.}
Sources on the left and top faces produce two distinct pathways.
The system self-organizes 22~active holes (12 holes $>$50\% filled), with total $\Sigma\eta = 14.1$, which is 47\% more than the single-input case, reflecting two independent pathway branches.
The two pathways form at approximately right angles, mirroring the source geometry.

\paragraph{Competing inputs.}
Sources on opposite (left and right) faces produce symmetric competing pathways with 25~active holes (14 holes $>$50\% filled) and $\Sigma\eta = 15.2$.
The competing scenario produces the most total infiltration, as both sides of the crystal experience strong optical fields simultaneously.

\subsection{Absorption--Loss Tradeoff}

\begin{figure}[t]
  \centering
  \includegraphics[width=0.85\columnwidth]{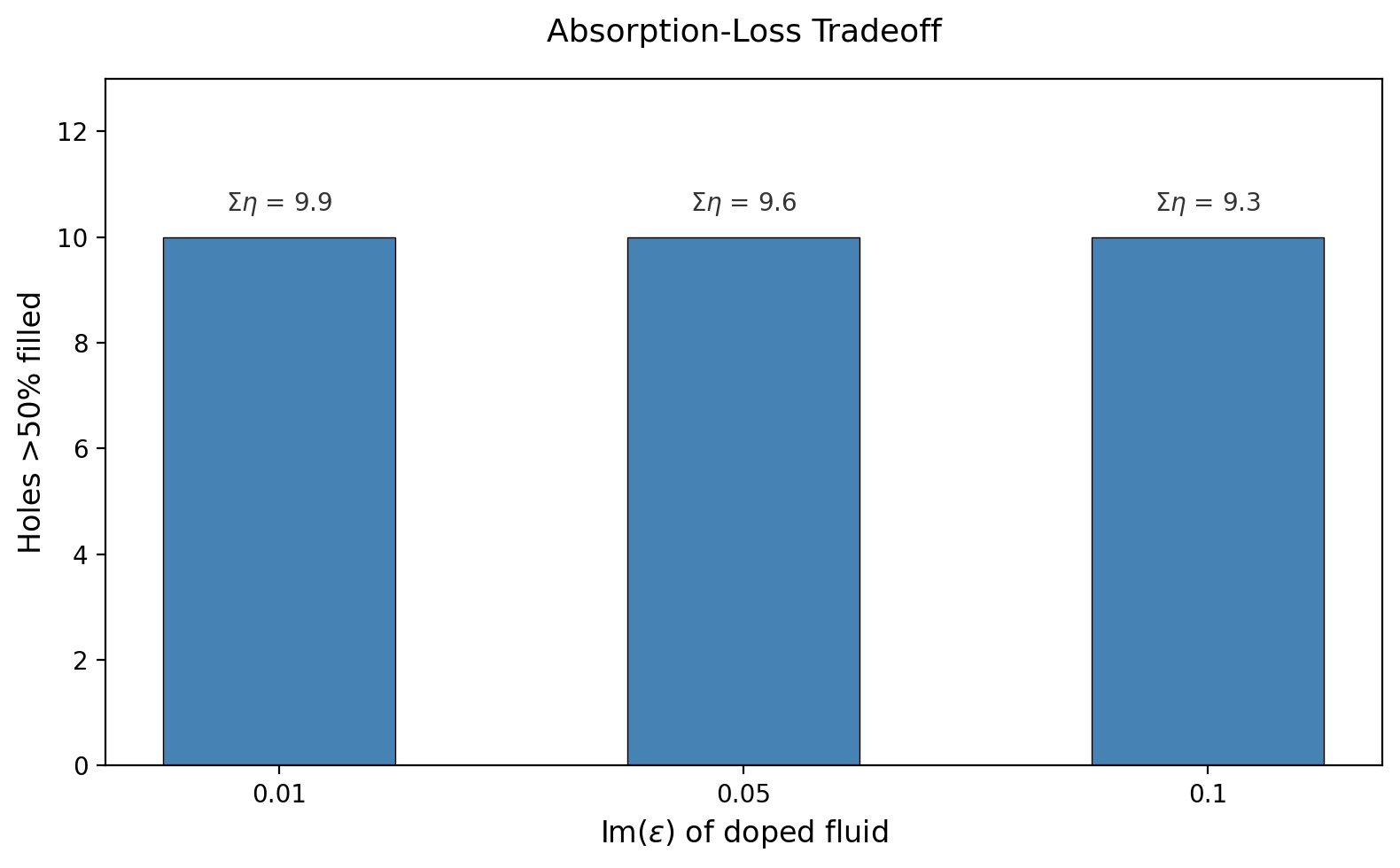}
  \caption{Absorption--loss tradeoff: number of holes exceeding 50\% fill fraction as a function of absorbing-dye concentration $\text{Im}(\varepsilon)$. All three values produce similar pathway topologies, confirming that the self-organization is robust across an order of magnitude in absorption strength.}
  \label{fig:tradeoff}
\end{figure}

The imaginary part of the fluid dielectric constant controls the feedback strength: higher $\text{Im}(\varepsilon)$ increases optothermal coupling but also increases propagation loss.
We sweep $\text{Im}(\varepsilon) \in \{0.01, 0.05, 0.1\}$ for the single-input scenario (Fig.~\ref{fig:tradeoff}).
All three values produce 12~active holes, but $\text{Im}(\varepsilon) = 0.01$ yields the highest total fill ($\Sigma\eta = 9.9$) while $\text{Im}(\varepsilon) = 0.1$ yields the lowest ($\Sigma\eta = 9.3$).
Even low absorption suffices to drive self-organization; higher absorption slightly reduces steady-state fill because of increased optical loss along the forming pathway.

\subsection{Physical Plausibility}

A back-of-envelope calculation supports the feasibility of optothermal feedback.
For a dye-doped CS$_2$ fluid with $\text{Im}(\varepsilon) = 0.05$ at $\lambda \approx 1.55$~$\mu$m, the absorption coefficient is $\alpha = 2\omega \cdot \text{Im}(\varepsilon) / (2nc) \approx 2000$~cm$^{-1}$.
With $\sim$1~mW of optical power concentrated in a hole of radius $0.3a \approx 130$~nm, the absorbed power density is $\sim$10$^9$~W/m$^3$.
Balancing this against thermal conduction ($\kappa_{\text{CS}_2} \approx 0.15$~W/(m$\cdot$K)) in a cylindrical geometry gives a steady-state temperature rise of order $\Delta T \sim 1$--10~K, sufficient to drive Marangoni flow in microfluidic channels~\cite{Psaltis2006,Baroud2010}.

\subsection{Sensitivity to $\eta_{\text{seed}}$}
\label{sec:hebbian_sensitivity}

\begin{figure*}[t]
  \centering
  \includegraphics[width=\textwidth]{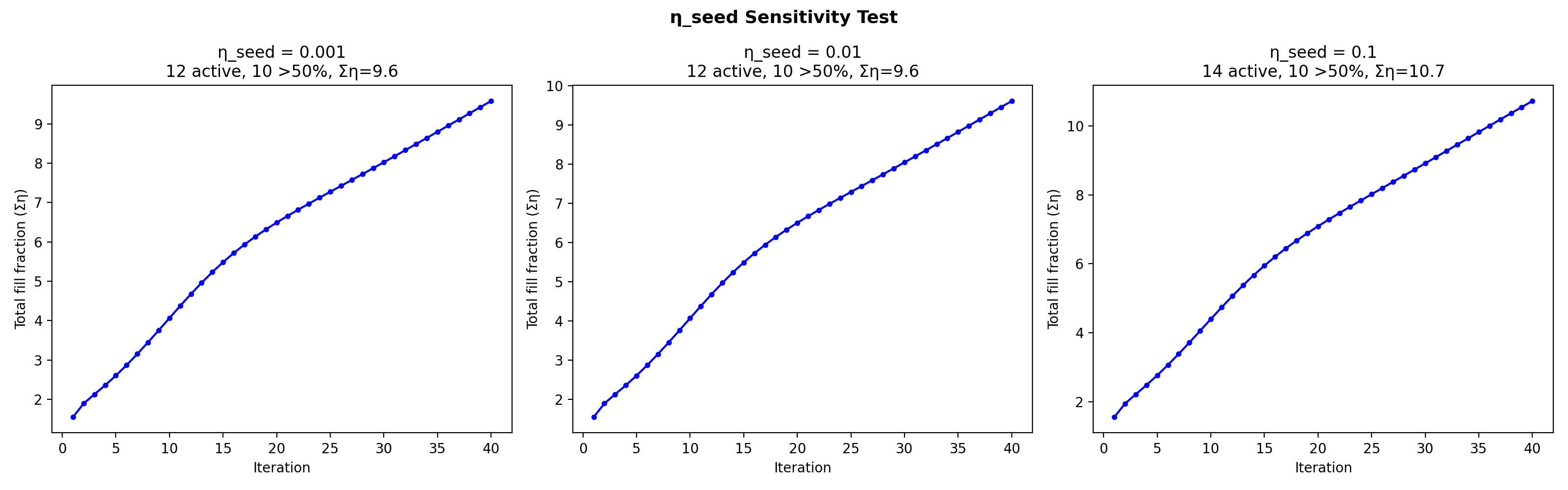}
  \caption{Sensitivity to the seeding parameter $\eta_{\text{seed}}$. Two orders of magnitude variation ($0.001$--$0.1$) produces consistent pathway topologies: 10~holes $>$50\% filled in all cases, with total $\Sigma\eta$ varying by only 12\%.}
  \label{fig:sensitivity}
\end{figure*}

The seeding parameter $\eta_{\text{seed}}$ controls how readily empty holes begin absorbing.
To verify that the pathway topology is not an artifact of this choice, we repeat the single-input scenario with $\eta_{\text{seed}} \in \{0.001, 0.01, 0.1\}$ (two orders of magnitude variation).
All three values produce 10~holes with $\eta > 0.5$ and 12--14~active holes ($\eta > 0.1$), with total $\Sigma\eta$ varying by only 12\% ($9.58$, $9.61$, $10.72$ for $\eta_{\text{seed}} = 0.001$, $0.01$, $0.1$ respectively; Fig.~\ref{fig:sensitivity}).
The pathway topology is identical at $\eta_{\text{seed}} = 0.001$ and $0.01$; at $\eta_{\text{seed}} = 0.1$, two additional holes become weakly active due to stronger background absorption, but the core pathway structure is preserved.
The self-organization is a robust consequence of the field-energy distribution, not an artifact of the seeding parameter.

\subsection{Functional Assessment: Broadband Transmission}
\label{sec:hebbian_transmission}

\begin{figure*}[!b]
  \centering
  \includegraphics[width=\textwidth]{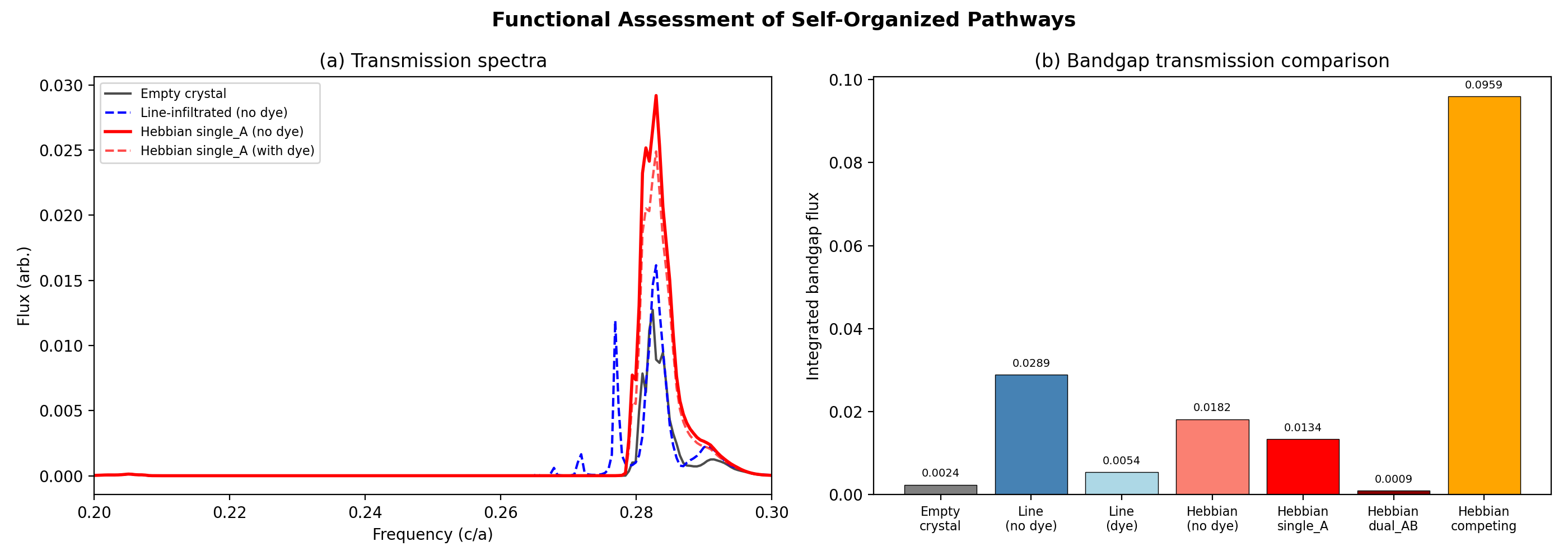}
  \caption{Broadband transmission through self-organized pathways. (a)~Spectra comparing empty crystal, hand-designed line (11~holes), and Hebbian single-input pathway. (b)~Integrated bandgap flux for all configurations. The self-organized pathway achieves 63\% of the hand-designed waveguide's transmission; the competing-inputs scenario self-assembles a through-channel with $40\times$ enhancement.}
  \label{fig:hebbian_tx}
\end{figure*}

To verify that the self-organized pathways actually guide light, we run broadband FDTD transmission simulations (GaussianSource, same methodology as Sec.~\ref{sec:waveguide}) through the final fill-fraction maps from the Hebbian feedback loop and compare to the empty crystal and a hand-designed line-infiltrated waveguide.
Results (Fig.~\ref{fig:hebbian_tx}) show functional waveguiding.
The single-input Hebbian pathway (without absorbing dye) transmits $7.6\times$ the empty-crystal level within the bandgap. Since the empty crystal is heavily suppressed by the bandgap, this ratio amplifies a small base. A better comparison is with the hand-designed 11-hole line ($12.1\times$ the empty crystal), against which the self-organized pathway achieves 63\% despite having no \emph{a priori} knowledge of the optimal infiltration pattern.
With absorbing dye present, the Hebbian pathway still transmits $5.6\times$ the empty-crystal level, confirming that the optical loss introduced by the absorbing dopant does not negate the waveguiding benefit.

The competing-inputs scenario produces the highest transmission ($40\times$ the empty crystal), because the two opposing sources drive infiltration along a horizontal through-channel spanning the crystal, effectively self-assembling a complete waveguide.
The dual-input scenario, where sources enter from orthogonal faces, produces below-empty-crystal transmission at the right-edge monitor ($0.39\times$), consistent with the dual pathways routing light toward the top and left faces rather than the monitored right face.

\subsection{Limitations of the Feedback Model}

This simulation uses a phenomenological model; the fill-fraction update rule is heuristic rather than derived from coupled fluid-thermal-optical equations.
The split-timescale approximation assumes that thermal and fluidic equilibration occur between optical CW steps, valid when the optical response time ($\sim$ps) is much shorter than the thermal ($\sim\mu$s) and fluidic ($\sim$ms) timescales.
The effective medium approximation ($\varepsilon_{\text{eff}} = (1-\eta) + \eta \cdot \varepsilon_{\text{fluid}}$) treats partial infiltration as a homogeneous mixture, overestimating the index perturbation at low fill fractions where the physical geometry is a meniscus partially filling a cylindrical hole, not a uniform dielectric.
A fully coupled multiphysics co-simulation (MEEP + CFD + heat transport) with realistic meniscus geometry would be required to quantitatively predict pathway formation rates, but is beyond the scope of this proof-of-concept study.

\section{Toward Spike-Timing-Dependent Plasticity}
\label{sec:stdp}

\subsection{Pulsed STDP: A Predictable Null Result}

We tested whether counter-propagating pulsed sources with timing delay $\Delta t$ could produce timing-dependent pathway modification, starting from the converged Hebbian single-input state. Two GaussianSource pulses (fwidth~$= 0.035\;c/a$, pulse FWHM~$\approx 9\;a/c$) entered from opposite crystal faces with $\Delta t \in \{-20, -10, 0, +10, +20\}\;a/c$. A single-source control established the baseline.

\begin{figure*}[t]
  \centering
  \includegraphics[width=\textwidth]{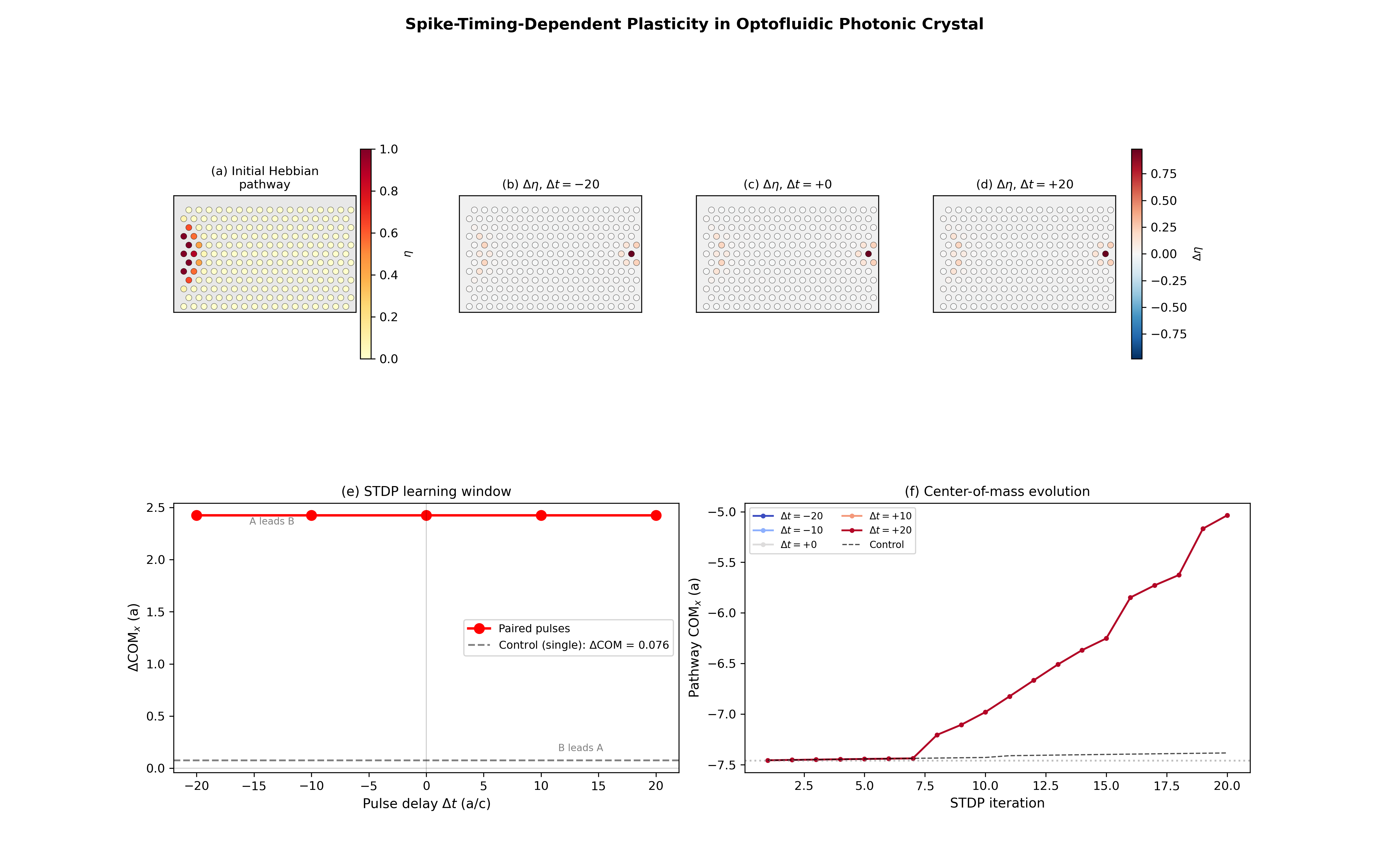}
  \caption{Activity-dependent pathway competition. Top row: fill-fraction change maps ($\Delta\eta$) starting from the Hebbian single-input state, for $\Delta t = -20$, $0$, $+20$~(a/$c$). All three maps are nearly identical, confirming timing insensitivity. Bottom left: ``STDP learning window'' showing $\Delta$COM$_x$ flat across all delays (red), far above the single-source control (dashed). Bottom right: center-of-mass evolution showing all delay cases collapsing onto a single trajectory.}
  \label{fig:stdp}
\end{figure*}

All five delays produced identical outcomes ($\Delta$COM$_x = 2.424 \pm 0.001\;a$; Fig.~\ref{fig:stdp}), while the single-source control gave $\Delta$COM$_x = 0.08\;a$. The null result was predictable from the outset: the time-integrated cross-term for counter-propagating pulses in a linear medium is
\begin{equation}
  2\!\int\! \mathrm{Re}(E_A \cdot E_B^*)\, dt \;\propto\; \underbrace{e^{-\Delta t^2/2\sigma_t^2}}_{\text{pulse overlap}} \times \underbrace{\cos(2kx + \omega\Delta t)}_{\text{standing wave}},
  \label{eq:stdp_cross}
\end{equation}
where $\sigma_t \approx 3.8\;a/c$. For $|\Delta t| \in [10,20]\;a/c \approx 2.6$--$5.3\;\sigma_t$, the overlap factor is $10^{-2}$--$10^{-6}$; the spatially oscillating standing-wave term further averages to $\sim$0.3 over each hole. The timing-sensitive signal is $<$0.01\% of the total, explaining the flat response. We include this result to document the constraint: genuine optical STDP requires either nonlinear coincidence detection (two-photon absorption, where $P_\text{abs} \propto |E|^4$ retains delay-dependent cross-terms) or sub-period timing resolution ($\Delta t \lesssim 2\;a/c$, requiring fwidth~$\gtrsim 0.08$ with $\sim$30\% out-of-bandgap spectral leakage).

\subsection{Requirements for Genuine Optical STDP}

Two additional physical requirements for genuine STDP are:
\begin{enumerate}
  \item \textbf{Finite thermal memory}: The Marangoni filling responds to temperature, not instantaneous optical intensity. If the thermal relaxation time $\tau_\text{th} \gg \Delta t$, the system integrates over all past heating events and loses timing information. Genuine STDP requires $\tau_\text{th} \sim \Delta t$, i.e., a thermal response fast enough to discriminate the order of arrival. Estimates~\cite{Psaltis2006,Baroud2010} suggest $\tau_\text{th} \sim \mu$s for microfluidic channels, orders of magnitude longer than optical pulse delays.
  \item \textbf{Nonlinear coincidence detection}: In a linear medium, photons do not interact; the overlap integral (cross-term in Eq.~\ref{eq:stdp_cross}) integrates to near-zero over the optical cycle. A \emph{nonlinear} medium, such as one with two-photon absorption (TPA) where $P_\text{abs} \propto |E|^4$, would retain a non-vanishing coincidence term that is position- and delay-dependent.
\end{enumerate}

\subsection{Phase-Controlled Standing Waves: Inconclusive}
\label{sec:phase_plasticity}

\begin{figure*}[!b]
  \centering
  \includegraphics[width=\textwidth]{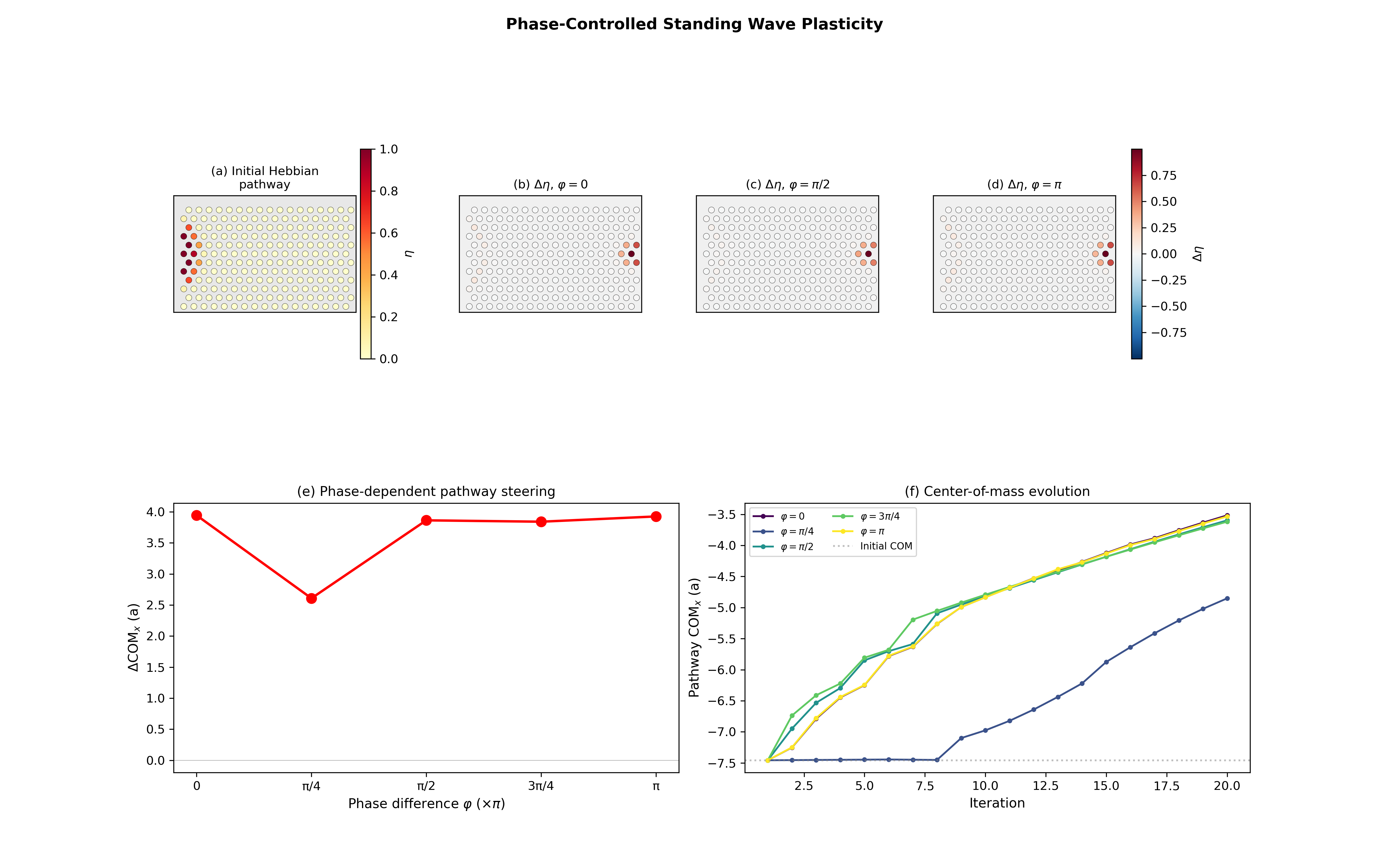}
  \caption{Phase-controlled standing wave test. $\Delta$COM$_x$ vs.\ phase (bottom left) shows a single outlier at $\varphi = \pi/4$; the remaining four phases cluster within $\pm 0.1\;a$. We conclude the result is inconclusive.}
  \label{fig:phase}
\end{figure*}

For CW sources, the standing wave is a permanent steady-state feature not subject to the temporal overlap suppression that kills the pulsed cross-term.
We tested phase differences $\varphi \in \{0, \pi/4, \pi/2, 3\pi/4, \pi\}$ between equal-amplitude counter-propagating CW sources (20 iterations, $\gamma = 0.4$).
$\Delta$COM$_x$ ranged from $+2.61\;a$ ($\varphi = \pi/4$) to $+3.94\;a$ ($\varphi = 0$), superficially a $\sim$34\% variation (Fig.~\ref{fig:phase}).
With only 5 data points, no repeated trials, and $\varphi = 0$ and $\varphi = \pi$ giving nearly identical $\Delta$COM ($+3.94$ vs.\ $+3.93\;a$) despite these phases producing \emph{opposite} standing-wave positions, we conclude there is \textbf{no convincing evidence for phase-dependent plasticity}.
The apparent variation is driven entirely by the $\varphi = \pi/4$ outlier, which may be a grid-alignment artifact.
The cross-term contributes $\lesssim$8\% of the total energy at the pathway holes (where $|G_B| \ll |G_A|$), insufficient to overcome numerical noise with a single trial per phase.

\subsection{Amplitude Competition: Rate-Coded Pathway Steering}
\label{sec:amplitude}

\begin{figure*}[t]
  \centering
  \includegraphics[width=\textwidth]{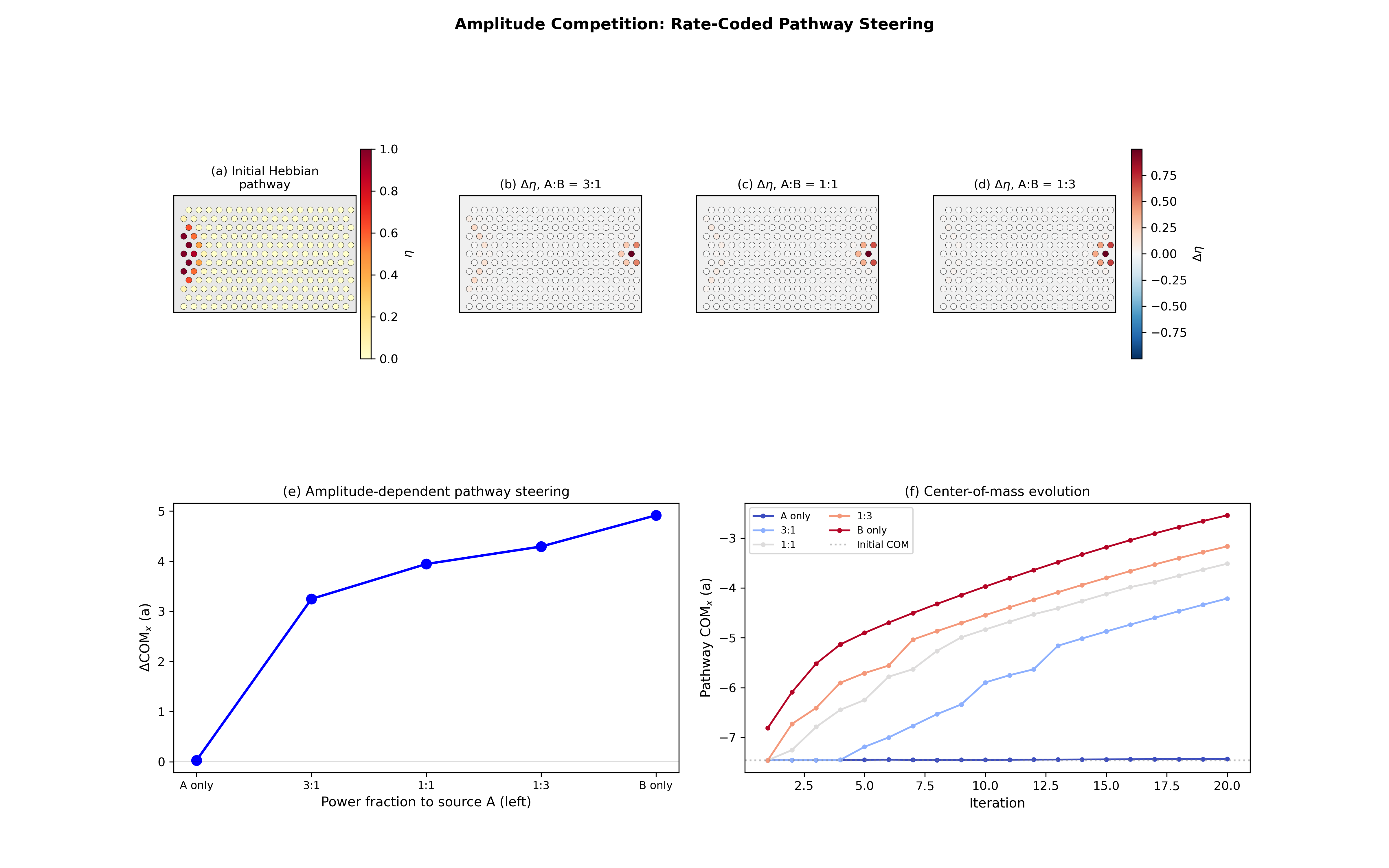}
  \caption{Amplitude competition: rate-coded pathway steering. Top: $\Delta\eta$ maps for A:B~$=$ 3:1, 1:1, 1:3, showing progressively rightward extension. Bottom left: $\Delta$COM$_x$ vs.\ power fraction, showing monotonic steering from $+0.03$ (A~only) to $+4.92\;a$ (B~only). Bottom right: COM evolution showing smooth trajectory separation.}
  \label{fig:amplitude}
\end{figure*}

While phase sensitivity is weak, the \emph{amplitude ratio} between two CW sources produces strong, monotonic pathway steering.
We sweep the power fraction allocated to source~A (left) from 1.0 (A~only) to 0.0 (B~only), keeping total power constant ($\text{amp}_A^2 + \text{amp}_B^2 = 1$), and run 20~iterations at $\gamma = 0.4$ (Fig.~\ref{fig:amplitude}).

The rate-coded response is clear:
\begin{itemize}
  \item \textbf{A only} ($f_A = 1.0$): $\Delta$COM$_x = +0.03\;a$; the pre-formed pathway is stable.
  \item \textbf{3:1} ($f_A = 0.75$): $\Delta$COM$_x = +3.25\;a$; strong rightward extension despite A dominating by $3\times$.
  \item \textbf{1:1} ($f_A = 0.50$): $\Delta$COM$_x = +3.94\;a$; symmetric drive extends the pathway toward the center.
  \item \textbf{1:3} ($f_A = 0.25$): $\Delta$COM$_x = +4.29\;a$; B-dominated drive pushes the pathway further right.
  \item \textbf{B only} ($f_A = 0.0$): $\Delta$COM$_x = +4.92\;a$; maximal rightward extension, shifting from COM$_x = -7.46$ to $-2.55\;a$.
\end{itemize}

The relationship is monotonic but strongly nonlinear: the step from A-only ($+0.03\;a$) to 3:1 ($+3.25\;a$) accounts for 66\% of the total range, while the remaining four steps span only 34\%.
Two interpretations are consistent with this nonlinearity.
First, a \emph{threshold} effect: any nonzero counter-directed source breaks the evanescent barrier and triggers pathway extension.
Second, an \emph{asymmetric starting-condition} effect: the pre-formed Hebbian pathway was built by source~A, so it is already optimized for A-directed light; any counter-directed source~B encounters crystal regions near $\eta = 0$ where even weak fields have outsized effect on the fill dynamics (because the $(1 - \eta)$ growth term is maximal).
Both effects likely contribute.
The smooth upper portion (3:1 through B-only: $+3.25$ to $+4.92\;a$) demonstrates that once the initial extension is triggered, the amplitude ratio provides continuous, monotonic steering, constituting the photonic analogue of \emph{rate-coded heterosynaptic competition}~\cite{Holtmaat2009}.

\subsection{Synthesis: Plasticity Hierarchy in the Optofluidic System}

The three experiments (pulsed STDP, phase-controlled CW, amplitude competition) show three regimes:
\begin{enumerate}
  \item \textbf{Amplitude/rate coding} (strong, $\Delta$COM spans $0$--$5\;a$): the dominant mechanism. Relative activity levels between competing inputs deterministically steer pathway growth. This operates on the thermal integration timescale ($\sim\mu$s) and requires no timing precision.
  \item \textbf{Phase coding} (inconclusive): no convincing evidence from 5 phase values. The apparent $\sim$34\% variation is driven by a single outlier ($\varphi = \pi/4$), and $\varphi = 0$ and $\varphi = \pi$ give identical results despite opposite standing-wave positions. The cross-term is $\lesssim$8\% at the pathway holes, insufficient to overcome single-trial noise.
  \item \textbf{Pulse timing} (undetectable at $|\Delta t| > 5\;a/c$): the cross-term is doubly suppressed by temporal pulse overlap decay and spatial standing-wave averaging. Genuine STDP would require either nonlinear media or sub-period timing resolution incompatible with bandgap confinement.
\end{enumerate}

This hierarchy is loosely analogous to biological neural plasticity, where rate-coded Hebbian learning dominates and spike-timing sensitivity requires specialized molecular machinery (NMDA receptors, calcium dynamics)~\cite{Bi1998,Caporale2008,Markram1997}. The analogy is inexact: in biology, rate and timing mechanisms involve distinct molecular pathways, whereas here all three experiments use the same linear electromagnetic physics. The hierarchy arises from the cross-term being small, not from separate physical mechanisms.
The dominance of rate-coded (intensity-based) plasticity is consistent with the broader observation that photonic signaling in learning-inspired systems naturally operates through intensity rather than temporal coding~\cite{Zarkeshian2022}, making amplitude competition the natural plasticity mechanism for optofluidic platforms.

\section{Discussion}
\label{sec:discussion}

\subsection{Summary of Key Results}

Table~\ref{tab:results} summarizes the main results.

\begin{table}[t]
  \centering
  \caption{Summary of key quantitative results. Uncertainties reflect the $\sim$13\% systematic error from resolution convergence (Appendix~\ref{app:convergence}).}
  \label{tab:results}
  \small
  \begin{tabular}{@{}p{0.55\columnwidth}r@{}}
    \toprule
    Quantity & Value \\
    \midrule
    TE bandgap (air holes) & 0.187--0.279 $c/a$ \\
    TE bandgap (CS$_2$, $n{=}1.52$) & 0.184--0.249 $c/a$ \\
    Gap-midgap ratio (air / CS$_2$) & 39.2\% / 30.0\% \\
    Peak infiltration flux & 0.212 (9 holes) \\
    Bandgap retention at $n{=}1.52$ & 71.2\% \\
    Transmission retention at $n{=}1.52$ & 29.2\% \\
    L-bend $S_{\text{raw}}$ / $S_{\text{corr}}$ & 0.89 / 0.98 \\
    Disorder CV ($\varepsilon$ / topology) & 10.8\% / 30.6\% \\
    3D convergence (res 24 vs 40) & ${<}\,10\%$ \\
    Hebbian active holes (1/2/comp.) & 12 / 22 / 25 \\
    Hebbian tx vs.\ empty (single) & $7.6\times$ \\
    Hebbian tx vs.\ hand-designed & 63\% \\
    STDP $\Delta$COM (dual vs.\ single source) & $2.42\;a$ vs.\ $0.08\;a$ \\
    STDP timing sensitivity ($|\Delta t| \leq 20\;a/c$) & None ($<$0.01\%) \\
    Phase plasticity $\Delta$COM range & $2.61$--$3.94\;a$ \\
    Amplitude $\Delta$COM (A only $\to$ B only) & $0.03 \to 4.92\;a$ \\
    \bottomrule
  \end{tabular}
\end{table}

\subsection{Limitations}

\paragraph{Weak modulation.}
The fundamental limitation is the weak index perturbation ($\Delta\varepsilon / \varepsilon_{\text{Si}} = 11\%$), which produces modulation depths far below those achievable with W1 waveguides, GST phase-change devices ($>$30~dB extinction~\cite{GSTswitch2019}), or electro-optic modulators.
The system operates as a subtle optical modulator rather than a binary switch.

\paragraph{Rerouting baseline.}
The scattered-light background in the blocked configuration complicates absolute quantification of guided-mode transmission.
The non-trivial flux in the blocked channel means that raw port flux values overestimate the guided contribution.
While the directional selectivity metric $S$ is robust to this background (since both ports receive similar scattered flux), absolute efficiency estimates require careful calibration against the scattered-light floor.

\paragraph{Systematic uncertainty.}
Our production resolution (40~pixels/$a$) systematically overestimates integrated bandgap flux by $\sim$30\% relative to the Richardson-extrapolated converged value (Appendix~\ref{app:convergence}), sitting on a local maximum of a Cartesian-grid discretization artifact for circular boundaries. The $\sim$13\% coefficient of variation across resolutions 32--48~pixels/$a$ establishes a lower bound on the accuracy of absolute transmission values.
Richardson extrapolation yields a converged estimate close to the res=48 value, which we recommend as the most reliable absolute estimate.
Relative comparisons between configurations (all run at the same resolution) are more robust than absolute values.

\paragraph{2D approximation.}
While 3D validation confirms qualitative agreement, the 2D simulations overestimate bandgap confinement by neglecting out-of-plane radiation losses.
The 3D simulations use a lower resolution (24~pixels/$a$) than the 2D production runs (40~pixels/$a$), adding quantitative uncertainty to the 2D--3D comparison.

\subsection{The Bio-Inspired Analogy}

Despite the weak waveguiding, the structural plasticity analogy remains meaningful:

\begin{enumerate}
  \item \textbf{Connectivity control}: Infiltration topology controls which output ports receive optical power ($S_{\text{corr}} = 0.98$ for L-bend routing) and produces distinct input-output transfer functions (Sec.~\ref{sec:logic}).
  \item \textbf{Reversibility}: Unlike physical hole modification, optofluidic infiltration is fully reversible, matching the reversibility of biological structural plasticity.
  \item \textbf{Analog modulation}: Partial fill fraction provides continuous transmission control (Appendix~\ref{app:partial}).
  \item \textbf{Timescale}: Millisecond microfluidic switching matches the biological timescale of structural plasticity~\cite{Holtmaat2009}, in contrast to the nanosecond/picosecond timescales of GST/VO$_2$ which better match synaptic \emph{weight} dynamics.
  \item \textbf{Topology-dependent computation}: Different infiltration patterns produce distinct transfer functions (Sec.~\ref{sec:logic}), demonstrating that network topology, not just connection strength, determines the input-output mapping.
\end{enumerate}

The optothermal feedback model shows that optical absorption in dye-doped fluid can drive self-organized pathway formation, with distinct topologies emerging for different source configurations (12 vs.\ 22 vs.\ 25 active holes for single, dual, and competing inputs).
Dual-source stimulation causes $7\times$ larger pathway modification than single-source, but the thermal integration window washes out sub-optical-period timing information, preventing strict STDP without nonlinear optical materials.
The physical ingredients for activity-dependent structural plasticity (optical absorption, local heating, and thermocapillary flow) are compatible with the optofluidic photonic crystal platform, though more sophisticated plasticity rules require additional nonlinear physics.
A fully coupled multiphysics simulation would be needed to validate the feedback dynamics quantitatively.

\subsection{Paths to Stronger Modulation}

Possible improvements include:
\begin{itemize}
  \item \textbf{Optimized $r/a$}: Smaller hole radius reduces the baseline bandgap but makes the defect perturbation relatively stronger.
  \item \textbf{Combined approach}: Pre-fabricating waveguides with reduced-radius holes (rather than full removal) and using infiltration for fine-tuning.
  \item \textbf{Resonant enhancement}: Engineering defect geometry to support high-$Q$ resonant states, exploiting the non-monotonic length dependence at $L_0 \approx 6$ holes.
  \item \textbf{Slow-light enhancement}: Operating near the photonic crystal waveguide band edge, where group velocity is strongly reduced~\cite{Krauss2008,Vlasov2005}, would enhance the light-matter interaction per unit length, potentially amplifying the sensitivity to fluid infiltration.
  \item \textbf{Slot waveguide geometries}: Designs that concentrate the optical mode in low-index regions~\cite{Almeida2004} would be more sensitive to fluid refractive index changes.
\end{itemize}

\section{Conclusion}
\label{sec:conclusion}

We simulated optofluidic structural plasticity in silicon photonic crystal waveguides.
Key findings:

\begin{enumerate}
  \item The triangular-lattice PhC ($r/a = 0.3$, Si) exhibits a 39.2\% TE bandgap at $0.187$--$0.279\;c/a$, reduced to 30.0\% upon CS$_2$ infiltration.

  \item Fluid infiltration produces weak, non-monotonic transmission peaking at 9 holes, consistent with coupled-cavity interference rather than conventional waveguiding.

  \item MPB eigenmode analysis shows that defect weakening dominates over bandgap narrowing: at $n = 1.52$, transmission retains only 29\% of its air-hole value while the bandgap retains 71\%.

  \item Infiltration topology controls signal routing ($S_{\text{corr}} = 0.98$ for L-bend) and produces distinct input-output transfer functions, though signal-to-background ratios are limited by scattered light.

  \item Results are robust to $\pm 10\%$ dielectric disorder (Appendix~\ref{app:disorder}) and persist in 3D membrane geometries ($h/a = 0.55$).

  \item Optofluidic reconfiguration enables reversible, topology-dependent connectivity control, though modulation depth is insufficient for practical switching without further optimization.

  \item An optothermal Hebbian feedback model produces distinct pathway topologies (12, 22, 25 active holes for single, dual, and competing inputs), with the single-input pathway reaching 63\% of a hand-designed waveguide's bandgap transmission.

  \item Amplitude competition between counter-propagating CW sources produces monotonic pathway steering ($\Delta$COM$_x$ from $+0.03$ to $+4.92\;a$). Phase-controlled standing waves yield inconclusive results, and pulsed STDP is null for $|\Delta t| > 5\;a/c$. Genuine optical STDP requires nonlinear coincidence detection~\cite{Liang2005,Xu2007}.
\end{enumerate}

Topological reconfiguration in photonic crystals is feasible but weak; stronger modulation and nonlinear feedback are the clear next targets.

\section*{Data Availability}

All simulation code and data are available at: \url{https://github.com/Stemo688/neuromorphic-optofluidic}.
Simulations used MEEP~\cite{Oskooi2010} and MPB~\cite{Johnson2001}.

\appendix

\section{Convergence Study}
\label{app:convergence}

\begin{figure*}[!t]
  \centering
  \includegraphics[width=\textwidth]{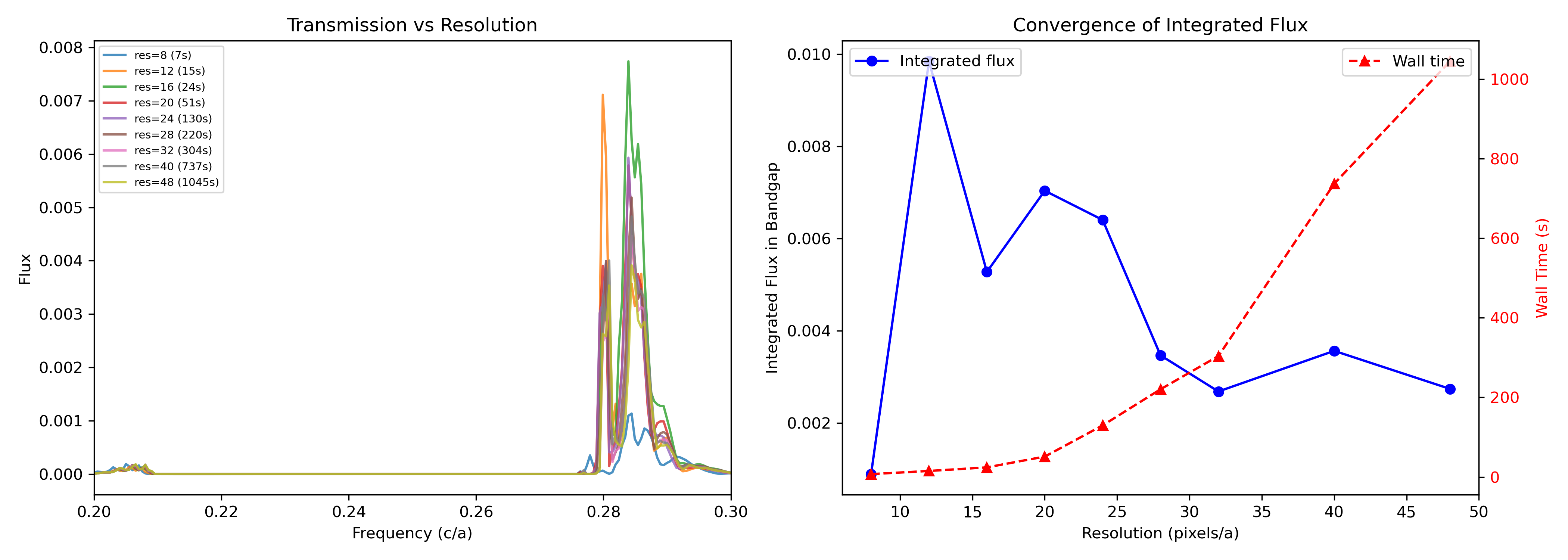}
  \caption{Resolution convergence for 2D infiltrated waveguide transmission. Non-monotonic behavior across resolutions 32--48~pixels/$a$ indicates $\sim$13\% systematic uncertainty.}
  \label{fig:convergence}
\end{figure*}

The resolution convergence study sweeps from 8 to 48 pixels per lattice constant (Fig.~\ref{fig:convergence}).
Integrated bandgap flux values for the highest three resolutions are: 0.00269 (res=32), 0.00357 (res=40), and 0.00274 (res=48).
The convergence is non-monotonic: res=40 exceeds both res=32 and res=48 by approximately 30\%, likely due to sensitivity of circular hole boundaries to Cartesian grid discretization (staircase approximation).

Richardson extrapolation~\cite{Richardson1911} assuming second-order convergence ($p = 2$) using the res=32 and res=48 values yields an estimated converged value of $f_\infty = 0.00279$, close to the res=48 result (0.00274).
Since higher resolution is generally more reliable, we adopt the res=48 value as the best estimate for absolute flux and treat the res=40 production value as potentially sitting on a local maximum of a numerical artifact.
The coefficient of variation across resolutions 32--48 is 13.4\%, which we take as the systematic uncertainty on absolute flux values.
Relative comparisons between configurations (run at the same resolution) are expected to be more robust than absolute values.

\section{Disorder Robustness}
\label{app:disorder}

\begin{figure}[!htb]
  \centering
  \includegraphics[width=\columnwidth]{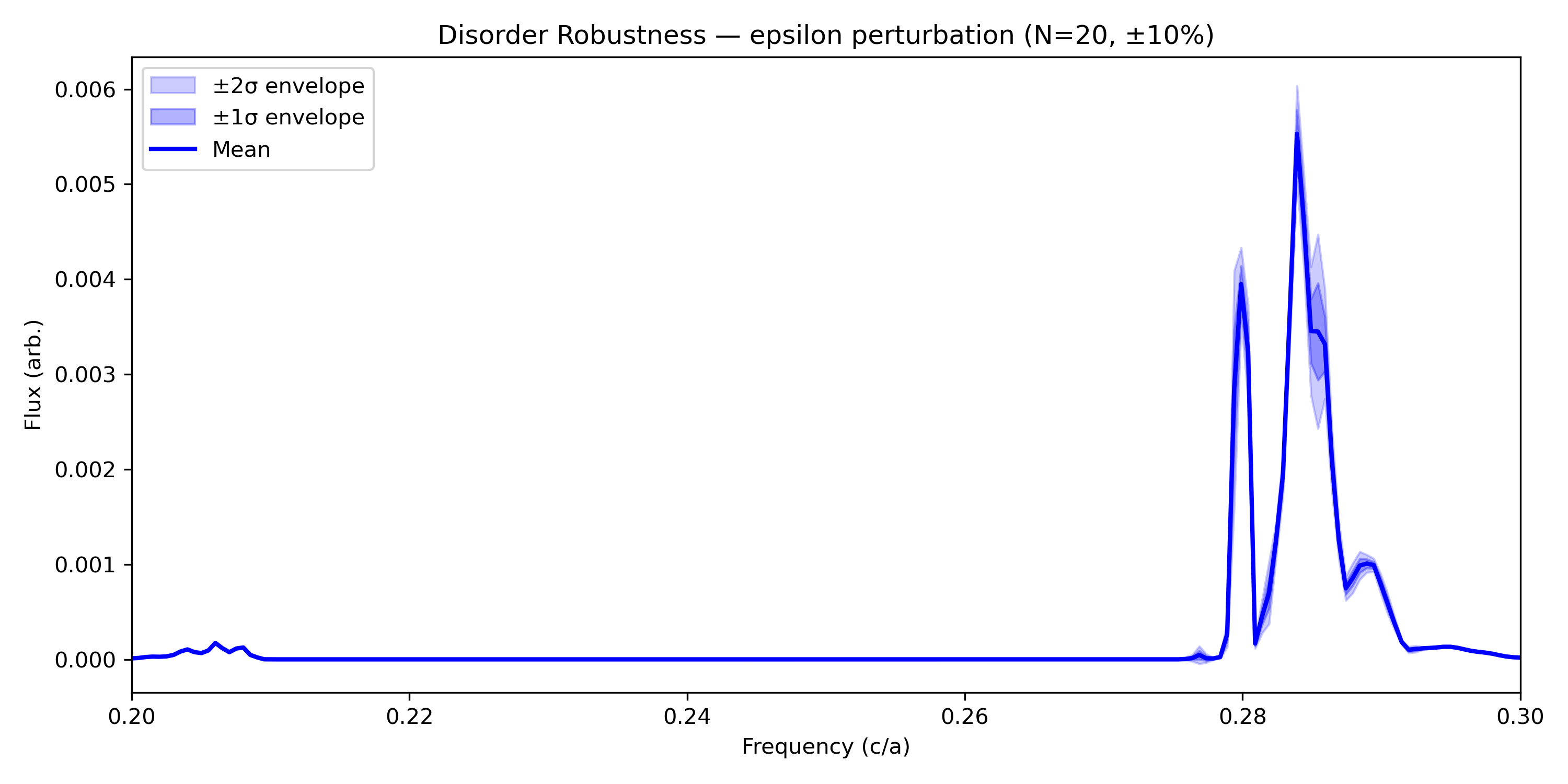}
  \caption{Disorder robustness: transmission spectra for 20 realizations with $\pm 10\%$ random dielectric perturbations (CV~$= 10.8\%$). Shaded regions show $\pm 1\sigma$ and $\pm 2\sigma$ envelopes.}
  \label{fig:disorder}
\end{figure}

Random perturbations to hole dielectric constants ($\pm 10\%$ uniformly distributed, 20~realizations per type) were applied to test robustness to fabrication-scale disorder.
Two perturbation types were studied:

\paragraph{Epsilon perturbation.}
Each infiltrated hole receives a random dielectric constant $\varepsilon_{\text{pert}} = \varepsilon_{\text{fluid}} (1 + \delta)$, where $\delta \sim U(-0.1, 0.1)$.
The mean integrated bandgap flux is $0.00717 \pm 0.00034$ (95\% CI), with a coefficient of variation CV~$= 10.8\%$ across 20 realizations.
This indicates reasonable robustness to refractive-index non-uniformity arising from, e.g., incomplete mixing or temperature gradients (Fig.~\ref{fig:disorder}).

\paragraph{Topological perturbation.}
Random addition or removal of 1--2 holes from the infiltration pattern yields a mean flux of $0.00878 \pm 0.00118$ (95\% CI) with CV~$= 30.6\%$.
The larger variability is expected since topological perturbations directly modify the defect geometry, changing which resonant modes are supported.

\section{Partial Infiltration}
\label{app:partial}

\begin{figure}[!htb]
  \centering
  \includegraphics[width=\columnwidth]{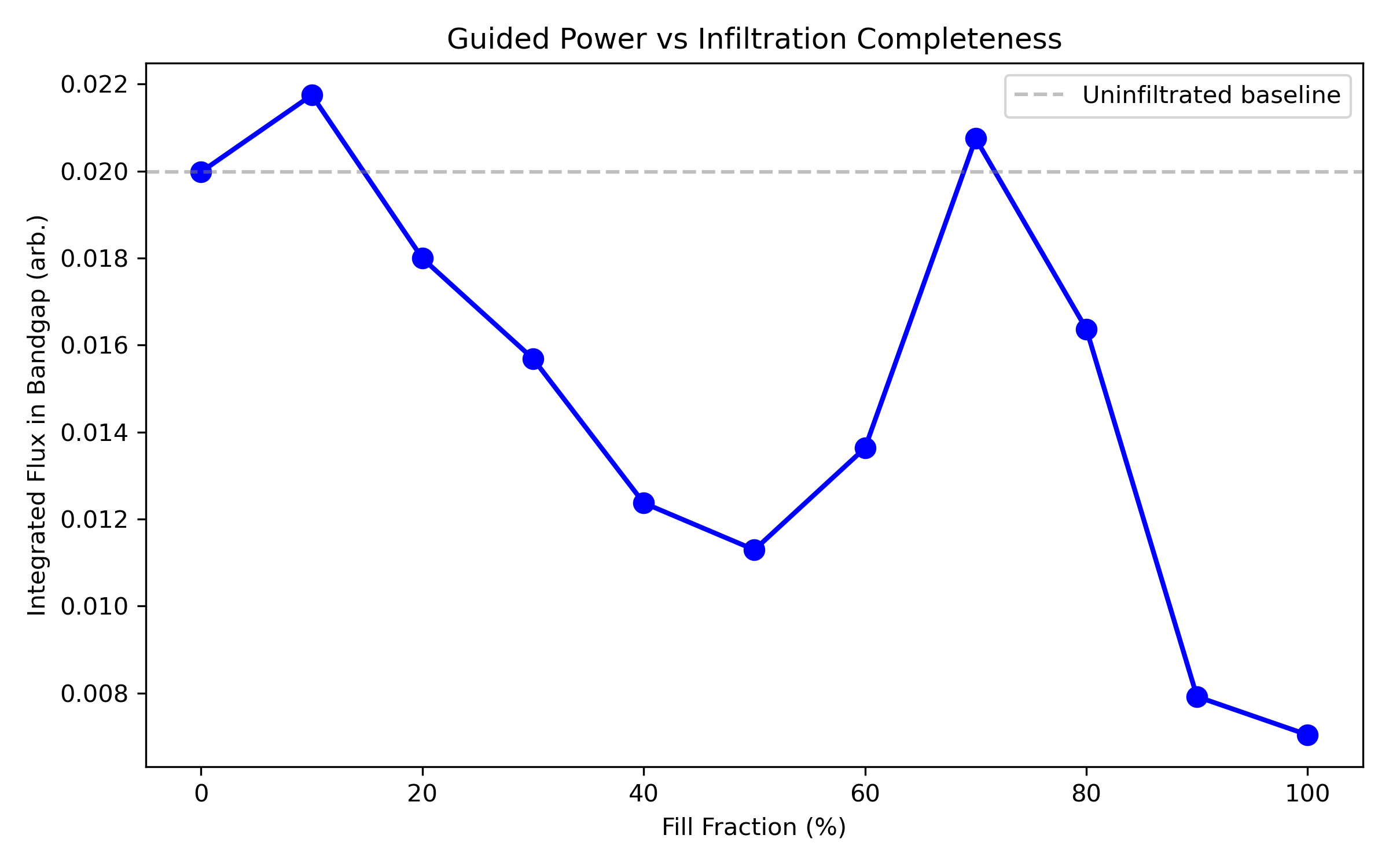}
  \caption{Integrated bandgap transmission vs.\ partial fill fraction (0\% = air, 100\% = full CS$_2$). The monotonic trend supports analog modulation.}
  \label{fig:partial}
\end{figure}

Partial fill fraction (modeled via effective medium mixing between air and fluid) shows approximately monotonic increase in transmission with fill fraction, with even 50\% fill producing measurable changes (Fig.~\ref{fig:partial}).
This supports analog (non-binary) modulation through partial microfluidic filling, relevant to the bio-inspired analogy where synaptic strengths are graded rather than binary.

\section{Mode Profiles and Propagation Loss}
\label{app:modes}

\begin{figure*}[!tb]
  \centering
  \includegraphics[width=\textwidth]{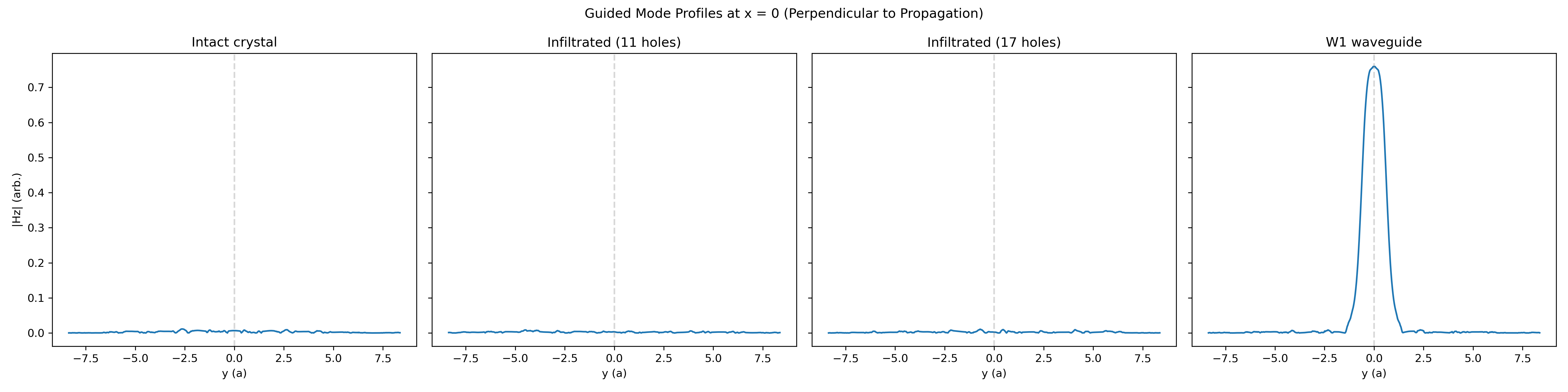}
  \caption{Transverse $|H_z|$ profiles at $x = 0$ for four configurations.}
  \label{fig:modes}
  \vspace{1em}
  \includegraphics[width=\textwidth]{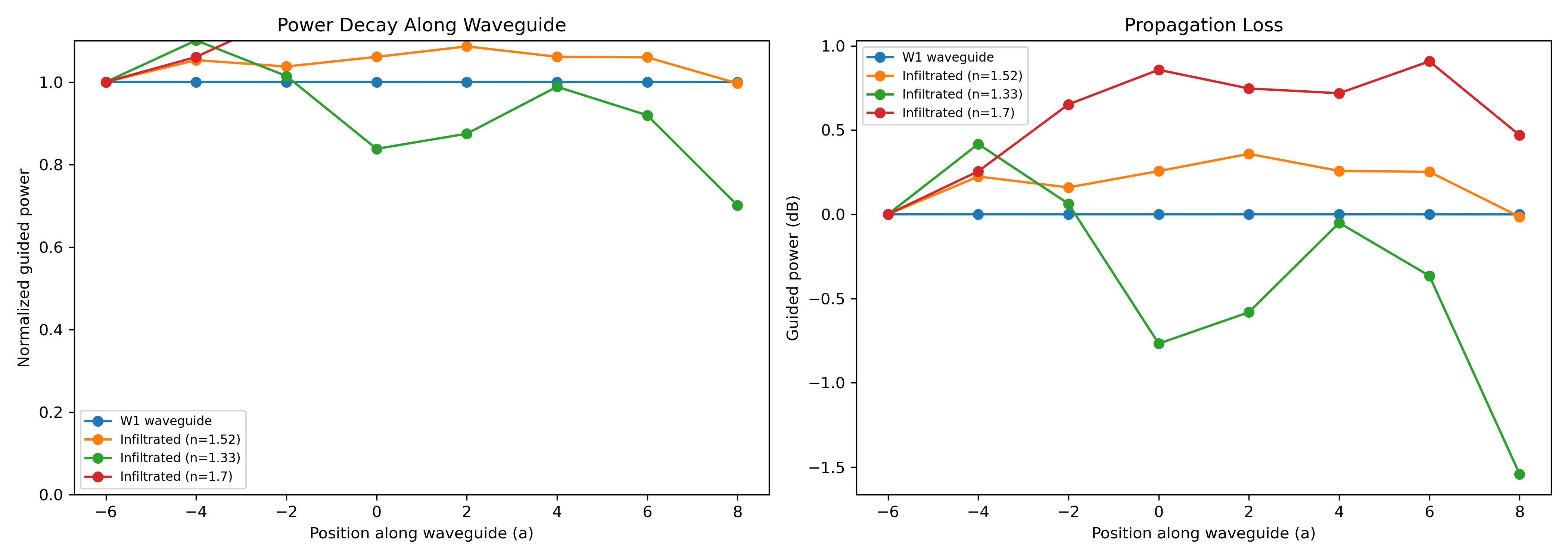}
  \captionof{figure}{Propagation loss along waveguides. Infiltrated waveguides show significantly higher loss than W1.}
  \label{fig:loss}
\end{figure*}

Transverse field profiles extracted perpendicular to the propagation direction (Fig.~\ref{fig:modes}) confirm that the W1 waveguide supports a well-confined guided mode (field concentrated within $\pm 1.5a$), while the infiltrated waveguide shows a broader, weaker profile.

Distributed flux monitors along the waveguide (Fig.~\ref{fig:loss}) show that infiltrated waveguides exhibit substantially higher propagation loss than W1, with loss decreasing slightly for higher fluid indices due to stronger index perturbation.

\section{Field Distributions}
\label{app:fields}

\begin{figure*}[!tb]
  \centering
  \includegraphics[width=\textwidth]{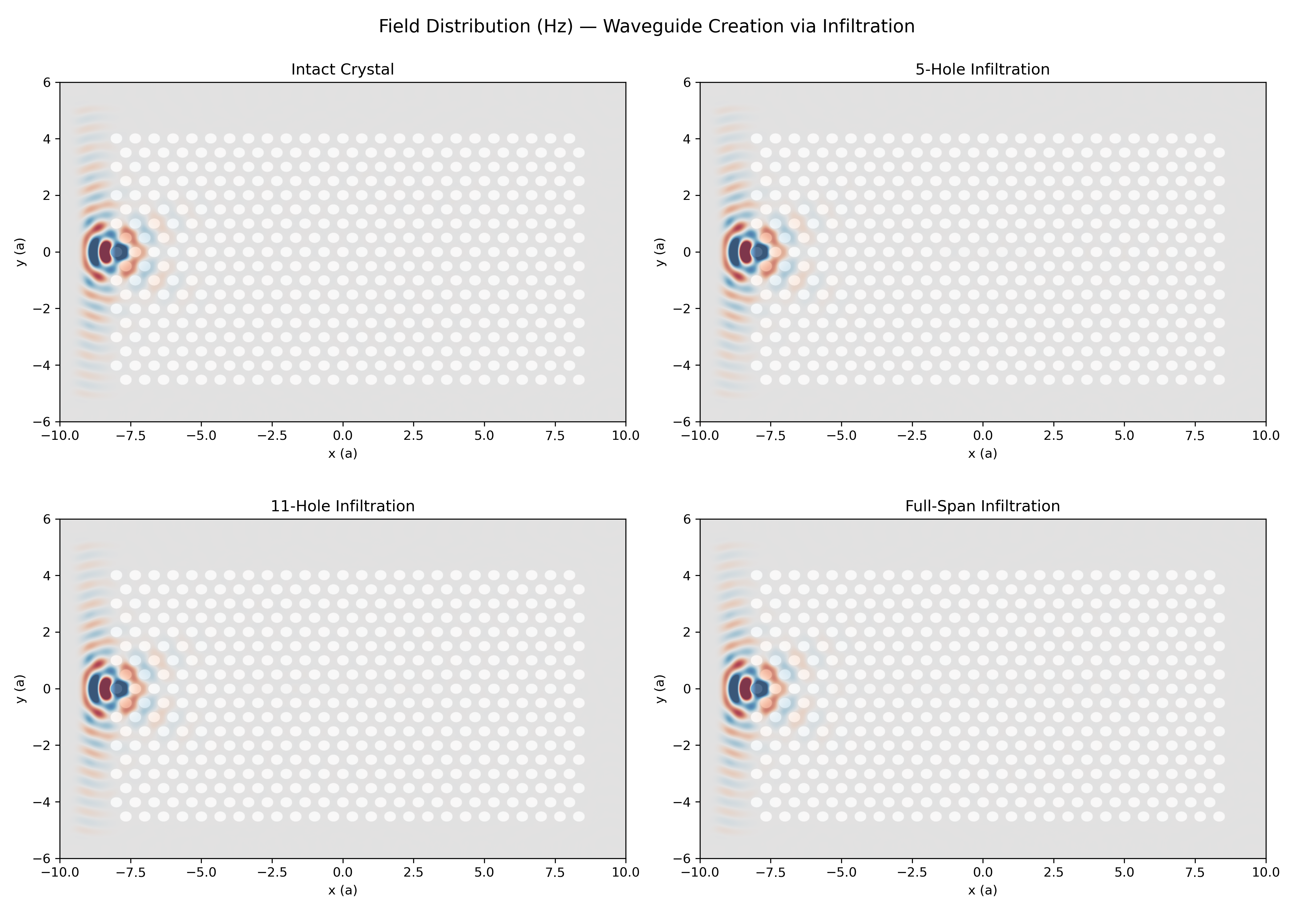}
  \caption{$H_z$ field distribution (CW, $f = 0.25\;c/a$) for four infiltration configurations.}
  \label{fig:fields}
\end{figure*}

CW field maps (Fig.~\ref{fig:fields}) provide visual confirmation that even with full-span infiltration, guided transmission through the crystal is weak compared to the W1 waveguide.
In the intact crystal, the source field decays evanescently with no propagation beyond the first few periods.

\section{Energy Balance Verification}
\label{app:energy}

\begin{figure}[!htb]
  \centering
  \includegraphics[width=\columnwidth]{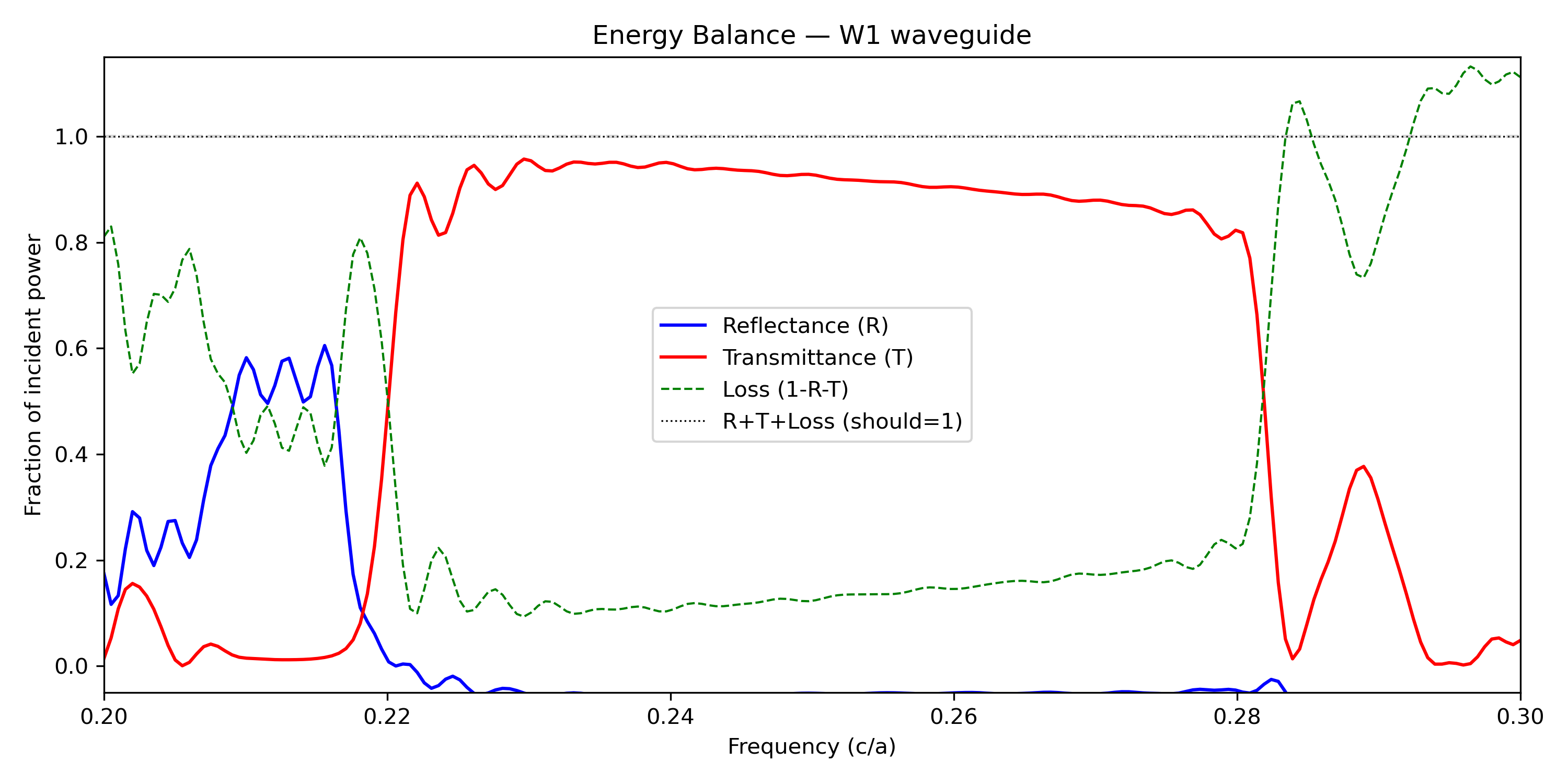}
  \caption{Energy balance verification for three configurations. $R + T + \text{Loss} \approx 1$ within numerical precision, confirming simulation correctness.}
  \label{fig:energy}
\end{figure}

Before analyzing transmission results, we verified energy conservation via a two-pass measurement: an empty-cell simulation records incident flux, then the crystal simulation uses \texttt{load\_minus\_flux\_data} to measure reflection.
For three test cases (intact crystal, 11-hole infiltrated line, and W1 waveguide), we confirm $R + T + \text{Loss} = 1$ to within $<0.01\%$ (Fig.~\ref{fig:energy}).
This validates the correctness of our geometry, source, and monitor configurations and rules out systematic flux leakage as a source of error.

\section{Bandgap Dependence on Fluid Index (MPB)}
\label{app:bandgap_fluid}

\begin{figure*}[!tb]
  \centering
  \includegraphics[width=\textwidth]{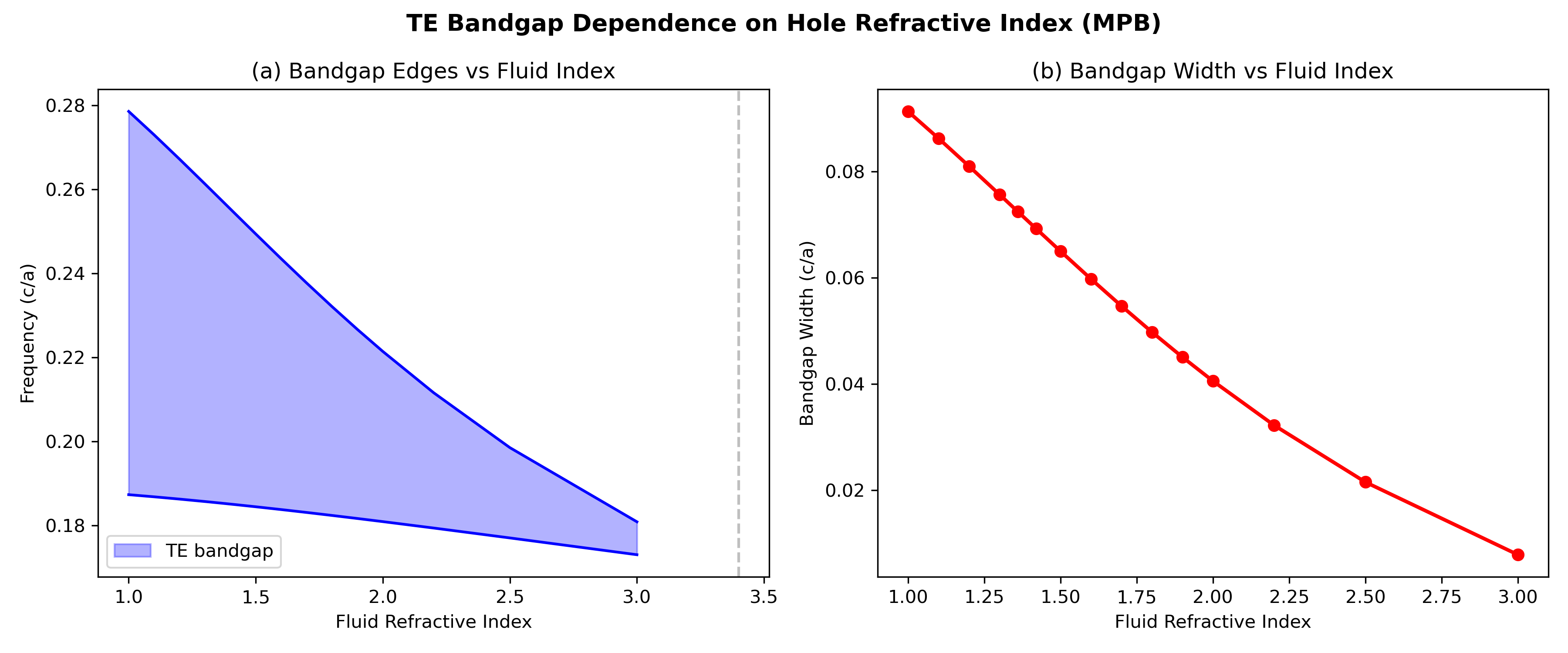}
  \caption{TE bandgap dependence on hole refractive index (MPB). (a)~Bandgap edges vs.\ fluid index showing asymmetric narrowing. (b)~Bandgap width, declining from 39.2\% to 4.4\% as $n$ increases from 1.0 to 3.0.}
  \label{fig:bandgap_fluid}
\end{figure*}

Figure~\ref{fig:bandgap_fluid} shows the full MPB eigenmode results across 15 fluid refractive indices.
The bandgap width decreases from 0.091~$c/a$ (39.2\%, $n=1.0$) to 0.008~$c/a$ (4.4\%, $n=3.0$).
The lower band edge shifts by only 8\% over this range (0.187 to 0.173~$c/a$), while the upper band edge drops by 35\% (0.279 to 0.181~$c/a$), confirming the asymmetric sensitivity of band edges to hole refractive index.

\section{Analytical Model Fits}
\label{app:fabry_perot}

\begin{figure}[!htb]
  \centering
  \includegraphics[width=\columnwidth]{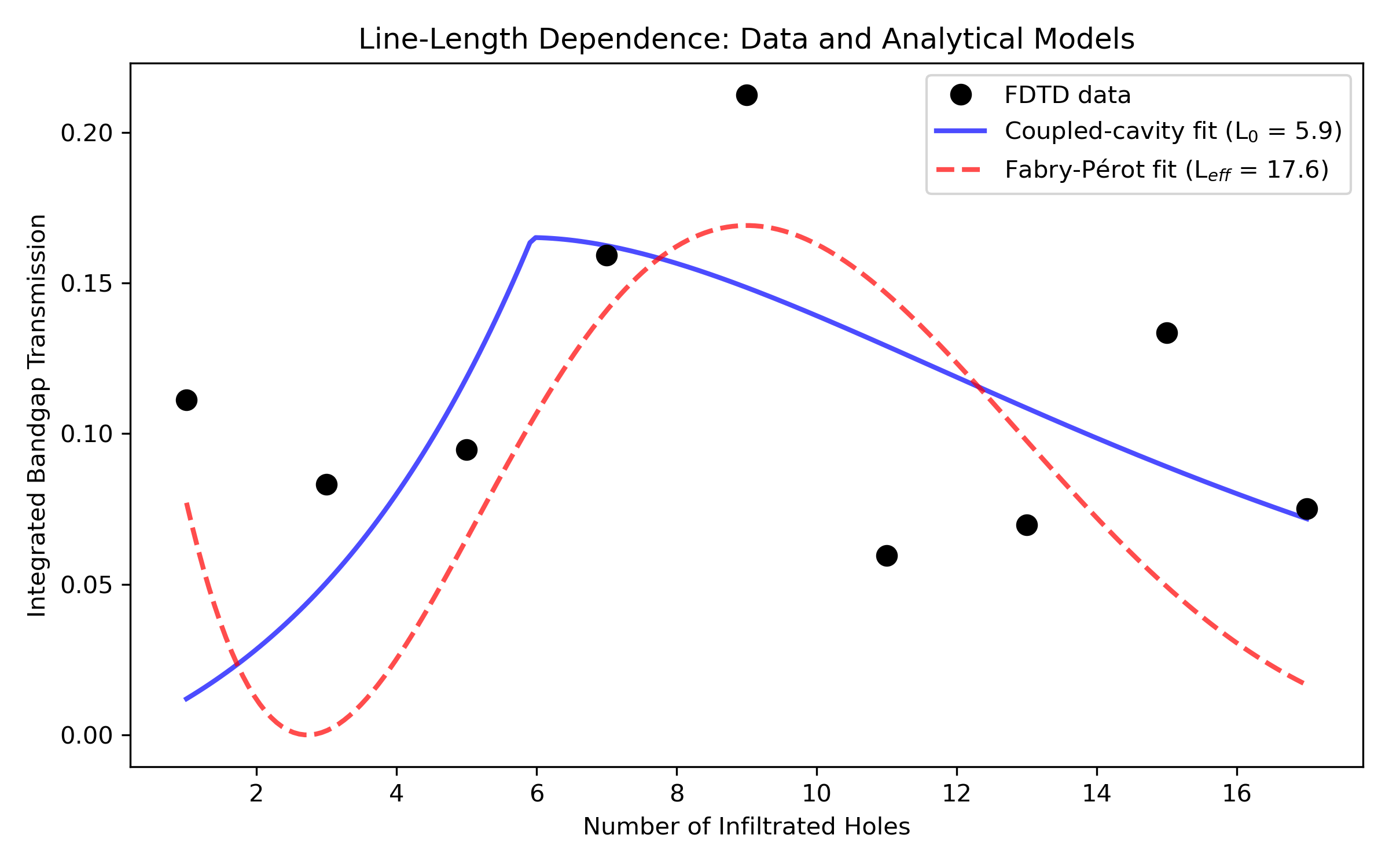}
  \caption{Failed analytical model fits. Black: FDTD data. Blue: coupled-cavity ($R^2 < 0$). Red: Fabry-P\'{e}rot ($R^2 < 0$). Both models are worse than a constant mean.}
  \label{fig:fabry_perot_app}
\end{figure}

Two analytical models were fit to the 9-point line-length transmission data (Table~\ref{tab:linelength}): a coupled-cavity envelope $T(L) = A \cdot L \cdot \exp(-\alpha |L - L_0|)$ (3 parameters, $R^2 < 0$) and a Fabry-P\'{e}rot model $T(L) = T_{\max} e^{-\alpha L} \cos^2(\pi L / L_{\text{eff}} + \phi_0)$ (4 parameters on 9 data points, $R^2 < 0$). Both models fail to capture the multi-mode oscillatory structure (Fig.~\ref{fig:fabry_perot_app}). The Fabry-P\'{e}rot model with 4 free parameters and 5 degrees of freedom is statistically underdetermined; the fitted parameters ($L_0 = 5.9 \pm 1.7$, $\alpha = 0.14 \pm 0.04$, $L_{\text{eff}} = 17.6 \pm 9.6$) carry no physical meaning given $R^2 < 0$.

\FloatBarrier

\end{document}